\documentclass[journal=nature, manuscript=article]{achemso}

\usepackage[version=3]{mhchem} 

\usepackage{amssymb}
\usepackage{xcolor}
\usepackage{soul}
\usepackage{siunitx}
\usepackage{nameref}
\usepackage[normalem]{ulem}
\usepackage{hyperref}

\usepackage{algorithm}
\usepackage{algpseudocode}
\usepackage{caption}
\usepackage{tikz}

\hypersetup{
    hidelinks=true,
	linkcolor=black,
	citecolor=black
}

\usepackage{xr}
\makeatletter
\newcommand*{\addFileDependency}[1]{
  \typeout{(#1)}
  \@addtofilelist{#1}
  \IfFileExists{#1}{}{\typeout{No file #1.}}
}
\makeatother
\newcommand*{\myexternaldocument}[1]{%
    \externaldocument{#1}%
    \addFileDependency{#1.tex}%
    \addFileDependency{#1.aux}%
}
\myexternaldocument{SI}

\author{F. Priante}
\affiliation{Department of Applied Physics, Aalto University, Helsinki FI-00076, Finland}
\author{N. Oinonen}
\affiliation{Department of Applied Physics, Aalto University, Helsinki FI-00076, Finland}
\author{Y. Tian}
\affiliation{International Center for Quantum Materials, Peking University, Beijing, 100871, China}
\author{D. Guan}
\affiliation{International Center for Quantum Materials, Peking University, Beijing, 100871, China}
\author{C. Xu}
\affiliation{Department of Applied Physics, Aalto University, Helsinki FI-00076, Finland}
\author{S. Cai}
\affiliation{Department of Applied Physics, Aalto University, Helsinki FI-00076, Finland}
\author{P. Liljeroth}
\affiliation{Department of Applied Physics, Aalto University, Helsinki FI-00076, Finland}
\email{peter.liljeroth@aalto.fi}
\author{Y. Jiang}
\affiliation{International Center for Quantum Materials, Peking University, Beijing, 100871, China}
\alsoaffiliation{Collaborative Innovation Center of Quantum Matter, Beijing, 100871, China}
\alsoaffiliation{CAS Center for Excellence in Topological Quantum Computation, University of Chinese Academy of Sciences, Beijing,China}
\alsoaffiliation{Interdisciplinary Institute of Light-Element Quantum Materials and Research Center for Light-Element Advanced Materials, Peking University, Beijing, 100871, China}
\email{yjiang@pku.edu.cn}
\author{A. S. Foster}
\alsoaffiliation{Department of Applied Physics, Aalto University, Helsinki FI-00076, Finland}
\affiliation{WPI Nano Life Science Institute (WPI-Nano LSI), Kanazawa University, Kakuma-machi, Kanazawa 920-1192, Japan}
\email{adam.foster@aalto.fi}

\date{\today}
\title{Structure discovery in Atomic Force Microscopy imaging of ice}

\begin{document}


\begin{abstract}

The interaction of water with surfaces is crucially important in a wide range of natural and technological settings. In particular, at low temperatures, unveiling the atomistic structure of adsorbed water clusters would provide valuable data for understanding the ice nucleation process. Using high-resolution Atomic Force Microscopy (AFM) and Scanning Tunneling Microscopy, several studies have demonstrated the presence of water pentamers, hexamers, heptamers (and of their combinations) on a variety of metallic surfaces \cite{Michaelides2007, Maier2016, Liriano2017, Dong2018}, as well the initial stages of 2D ice growth on an insulating surface \cite{Ma2020}. However, in all these cases, the observed structures were completely flat, providing a relatively straightforward path to interpretation. Here, we present high-resolution AFM measurements of several new water clusters on Cu(111), whose understanding presents significant challenges, due to both their highly 3D configuration and to their large size. For each of them, we use a combination of machine learning, atomistic modelling with neural network potentials and statistical sampling to propose an underlying atomic structure, finally comparing its AFM simulated images to the experimental ones. These results provide new insights into the early phases of ice formation, which is a ubiquitous phenomenon ranging from biology to astrophysics.
\end{abstract}

\section{Introduction}
Water-solid interfaces feature prominently in a wide spectrum of scientific and technological problems, encompassing material science, chemistry, biology and geology. A necessary prerequisite for their understanding is knowing how the water molecules will be structurally organized on the solid surface. The complex interplay between water-water and substrate-water interactions gives  rise to a highly diverse range of possible structures, forming one-dimensional \cite{Morgenstern1996, Yamada2006, Carrasco2009}, two-dimensional \cite{Lew2011, Forster2011, Ma2020} and three-dimensional \cite{Maier2016} configurations. 

Among this rich variety, ice nanoclusters \cite{Michaelides2007, Guo2014, Liriano2017, Dong2018} are of particular importance, as they enable sampling the vast space of metastable configurations explored by water molecules during the heterogeneous ice nucleation process. Atomically resolved images of these nanoclusters can be obtained from Atomic Force Microscopy (AFM) or Scanning Tunneling Microscopy (STM) experiments, under ultra-high vacuum and low temperature conditions, and using tip-functionalization \cite{Gross2009}. However, beyond clusters of only a few molecules, the resulting images are often difficult to analyze, due to the tendency of the nanoclusters to arrange in buckled, bilayered structures once enough water molecules have been deposited. The difficulty is further increased by the flexibility of their hydrogen bond framework, which, at close distances, can be significantly perturbed by an approaching microscope tip.

Due to these challenges, current investigations often focus on planar, monolayered nanoclusters, for which structural interpretation is relatively straightforward. To approach more general and three-dimensional cases, a promising route could be the application of recent advancements in machine learning (ML) for AFM image analysis. These techniques have already allowed to extract physical descriptors \cite{Alldritt2020}, electrostatic potential maps \cite{Oinonen2021}, ball-and-stick molecular representations \cite{carracedo-cosme2022} and molecular graphs \cite{Oinonen2022} from AFM image inputs. However, these models were trained on organic molecules in vacuum, which present significantly different chemical features compared to water molecules adsorbed on metal surfaces. Until architectures become available for robustly extrapolating across chemical space, it will remain necessary to generate additional data for applying automated AFM structure discovery to new domains. Indeed, by training on a custom water dataset, the two-dimensional configuration of waters in a Na\textsuperscript{+}4H\textsubscript{2}O hydrate was successfully predicted \cite{Tang2022}.

Furthermore, to fully reconcile an experimental observation with its predicted ice nanocluster structure, the underlying substrate must be taken into account. This usually implies carrying out a geometry relaxation of the hypothesized structure onto the surface, using quantum mechanical methods such as Density Functional Theory (DFT). For large and bilayered clusters, this is inherently difficult as very little information can be obtained with AFM about the organization of the lower ice layer. However, even for smaller monolayer cases, the irregular arrangement of water molecules means that even small variations in the initial hydrogen bond network and adsorption configuration on the substrate can generate drastically different final geometries. This holds true also for weakly reactive, hydrophobic metal surfaces, which are generally favored in experiments, as they don't cause further complications such as hydrogen dissociation. 

In this work, we tackle these challenges by developing a simulation workflow for structure discovery in high-resolution AFM imaging of large, buckled, mono and bi-layered ice nanoclusters on Au(111) and Cu(111) surfaces. We utilize the workflow on eight experimental AFM images, obtaining excellent agreement with the simulated AFM from the discovered atomic structures. Furthermore, we demonstrate robustness of the predictions upon their relaxation on surfaces, fully closing the loop between experiment and interpretation.

\section{Results}

\begin{figure}[h!] 
\begin{center}
\centering
\includegraphics[width=160mm]{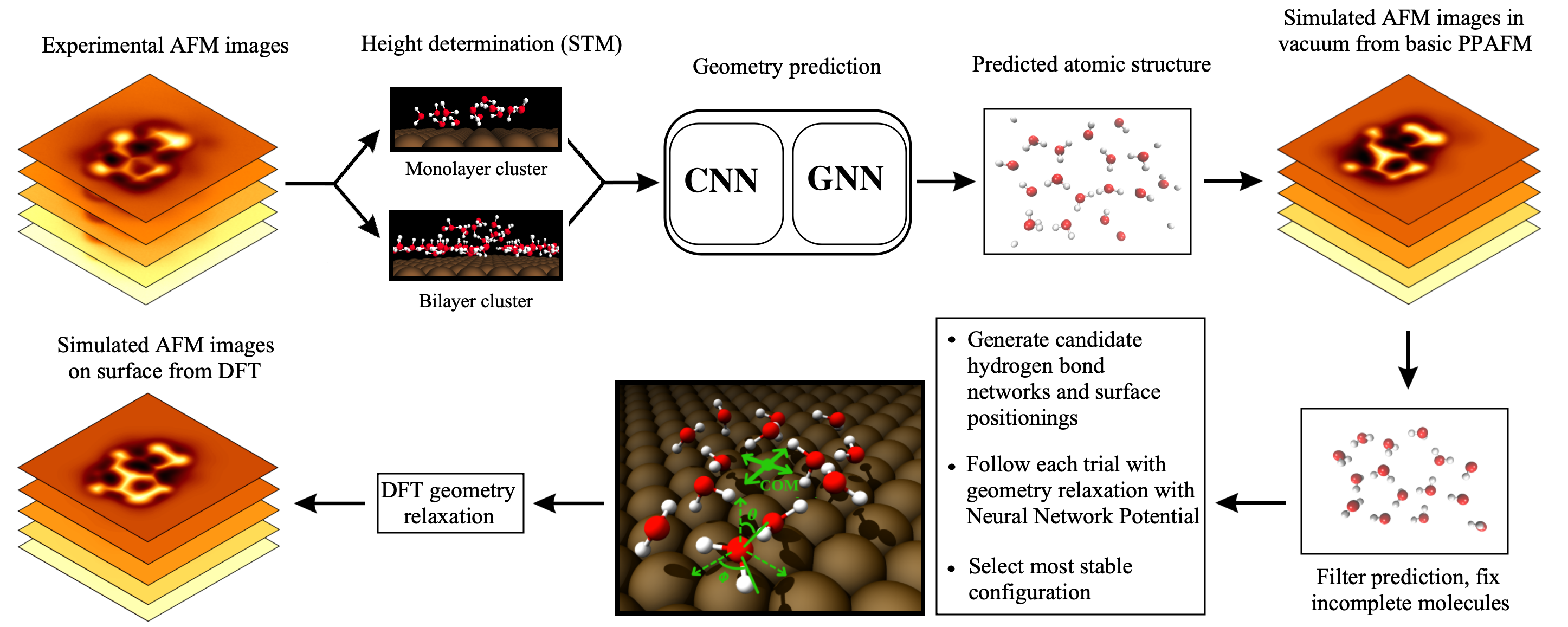}
\caption{Workflow for structure discovery in AFM images of ice nanoclusters. If Scanning Tunneling Microscopy (STM) is carried out in conjunction with the AFM, the monolayer or bilayer character of the nanocluster can be initially determined from the measured height difference to the substrate (see Fig.~\ref{fig:stm_profiles}). Then, an atomic structure prediction is inferred from the experimental AFM image stack. Incomplete molecules are fixed, while possibly unrealistic ones are eliminated. Next, multiple hydrogen bond networks and surface positions are rapidly evaluated by carrying out Neural Network Potential relaxations. The most stable resulting structure is further optimized with DFT, from which simulated AFM images can be obtained and compared to the initial experiment.}
\label{fig:workflow}
\end{center}
\end{figure}

The structure discovery workflow involves a sequence of steps summarized in Fig.~\ref{fig:workflow}. We apply it on seven experimental samples on a Au(111) surface, and on one sample on Cu(111). The full experimental AFM image sets are shown in Fig.~\ref{fig:afm_full} in the supplementary information (SI). In the following sections, we outline the individual components of the workflow in detail.

\subsection{Geometry prediction}

In the initial phase of the workflow, an AFM image stack is fed into our ML geometry prediction model, schematized in Fig.~\ref{fig:one-shot-graph}. Building upon our previous infrastructures \cite{Oinonen2021, Oinonen2022}, the model comprises an Attention U-Net convolutional neural network (CNN) for predicting atomic positions, and a Graph Neural Network (GNN) for identifying their corresponding atomic species.

Three different ML models were trained, covering each type of system we encountered in experiments. Particularly, we considered separately monolayer and bilayer nanoclusters on Au(111), and monolayers on Cu(111). The models were trained on distinct datasets of AFM images simulated with the Probe Particle Model (PPM) \cite{hapala2014}. The images were obtained using the Hartree potentials of randomly generated ice nanoclusters, previously relaxed with Neural Network Potentials (NNP), as shown in Fig.~\ref{fig:dataset_ml}. The NNPs themselves were also trained separately, one for each metal, using a diverse range of water structures, exemplified in Fig.~\ref{fig:dataset_nnp}. 

\begin{figure}[h!]
\begin{center}
\centering
\includegraphics[height=175mm]{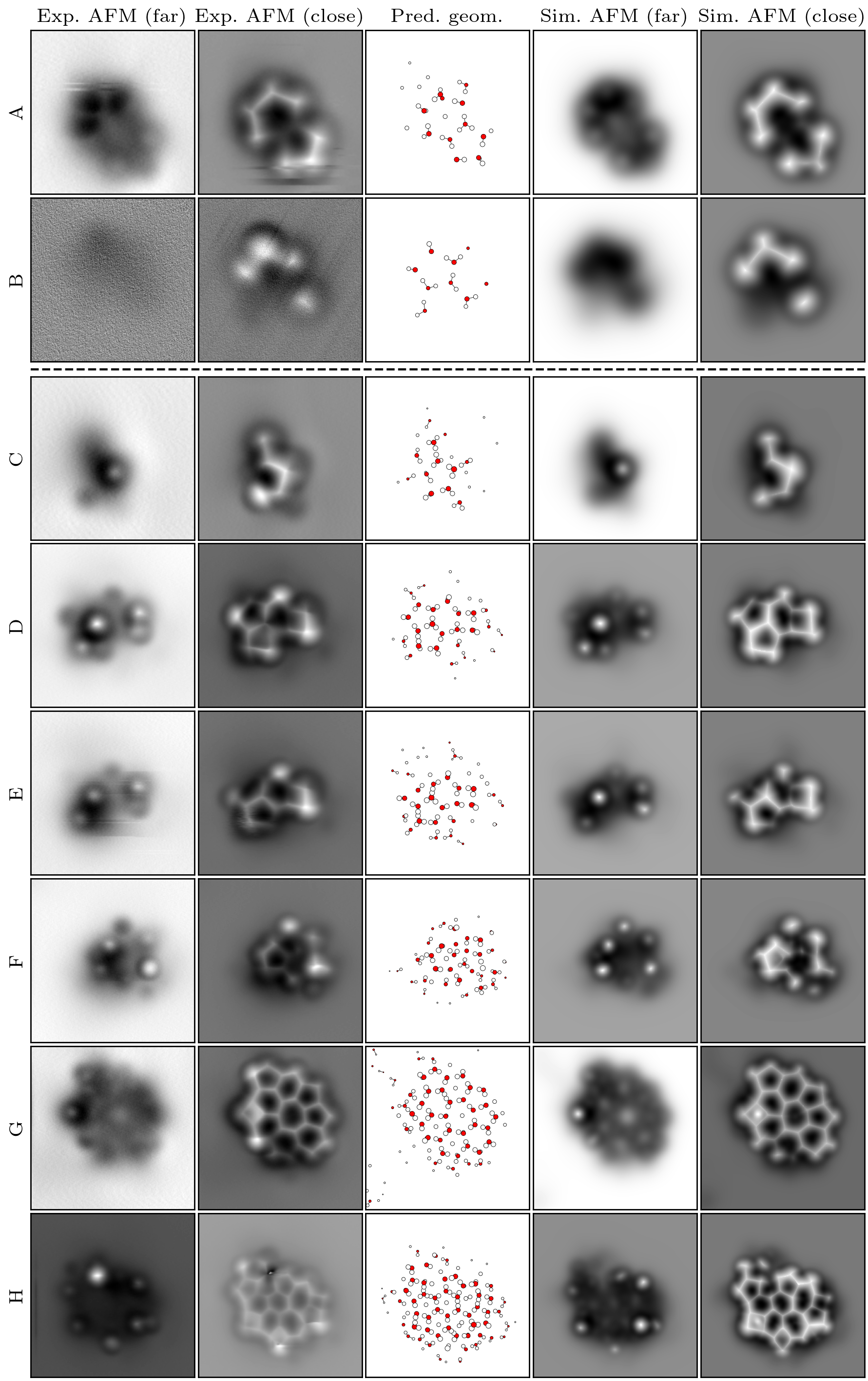}
\caption{Geometry predictions for experimental AFM images and corresponding simulated AFM images. Each row corresponds to one experiment. Experiments A and B are monolayer, and C-H are bilayer. On the left are the farthest and closest distance experimental AFM images, in the middle the predicted geometries, and on the right the simulations based on the predicted geometries. The tip-sample distances in the simulations are manually chosen in each case to visually match the experimental images. The sizes of the atoms in the geometry indicate the relative depths of the atoms.}
\label{fig:exp_comp}
\end{center}
\end{figure}

\begin{figure}[h]
	\centering
	\includegraphics[width=160mm]{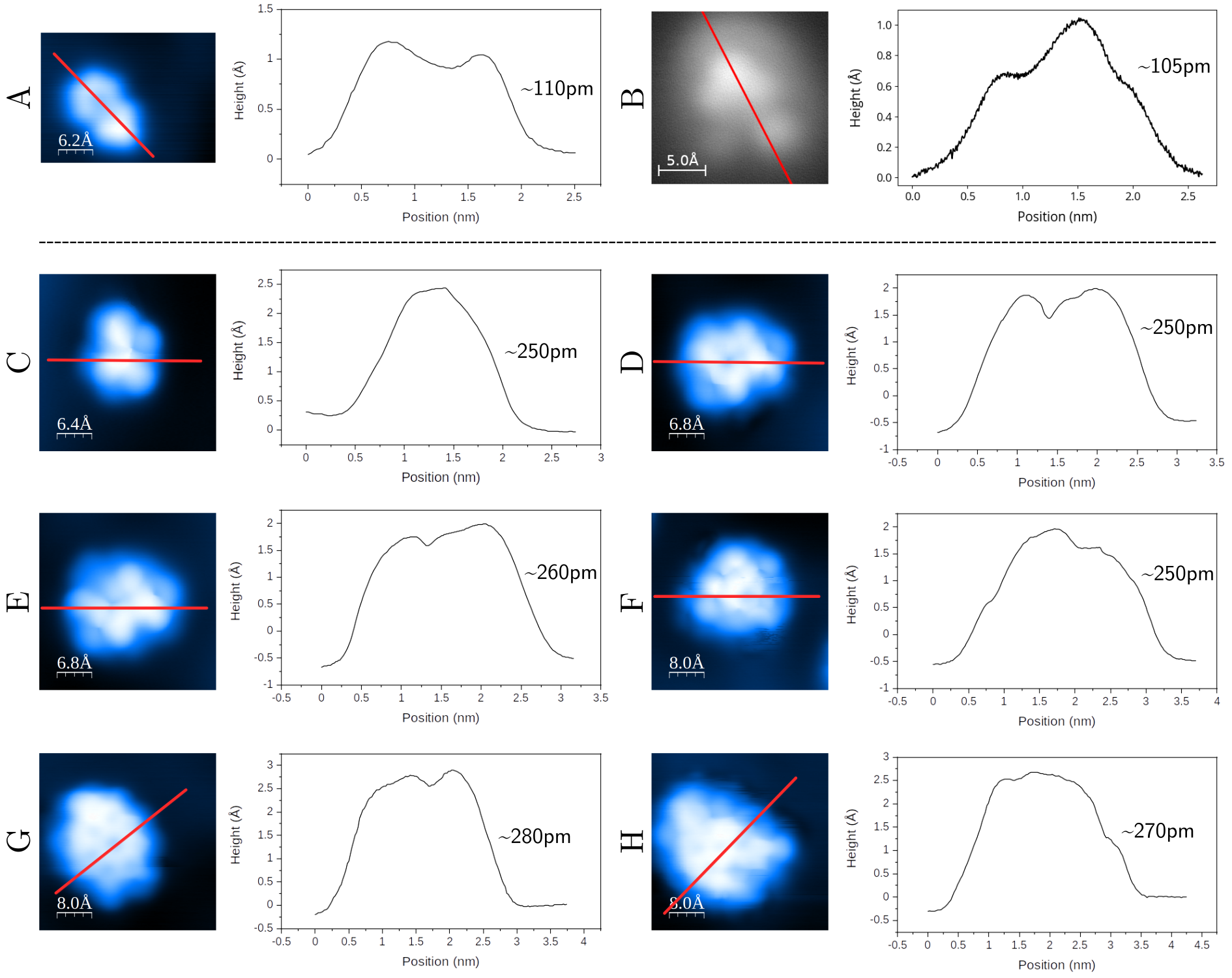}
	\caption{STM images and line profiles for each of the experiments. The height profile for A and B is significantly lower than for the other experiments, indicating that A and B are monolayer whereas the other ones are bilayer.}
	\label{fig:stm_profiles}
\end{figure}

The results are shown in Fig.~\ref{fig:exp_comp}, where we compare the experimental AFM images of the clusters, their predicted geometry, and the AFM simulations from the predictions. Of these experiments, two (A and B) are monolayer and others bilayer, which is indicated by different height profiles in STM line scans, shown in Fig.~\ref{fig:stm_profiles} in the SI. Sample B is on Cu(111) and the others on Au(111). For each sample, we include images at both high and low tip height, emphasizing the markedly three-dimensional character of these structures. We observe an excellent agreement between predictions and observations, with just a single very high atom in sample H which perhaps confused the model and was not accounted for. 

We found the performance of the geometry predictions to improve upon increasing the input AFM image stack size $n_z$, saturating at around $n_z = 10$. We carry out an in-depth analysis of this trend in Sec.\ "\nameref{sec:results_stack_size}" and Figs.~\ref{fig:loss_z}, \ref{fig:pred_z}, and \ref{fig:pred_zr} in the SI. In the same section we further show how, if using only a $n_z = 1$, then images closer to the sample provide more information, as expected.

\subsection{Prediction pruning}

In the least bright regions of the images, which are the hardest even for human experts to interpret, all our three models tend to generally predict the presence, rather than the absence of low-lying atoms. This is likely due to the ML model trying to reproduce as accurately as possible the experimental image features, which are generally sampled from a different distribution than the simulated ones the ML was trained on. This is a well-known problem when adapting models trained on synthetic data onto real-world scenarios, and the development of strategies to overcome it is still an active area of research \cite{Tobin2017, Taori2020, robust, Wang2022}. 

When moving on to the next step of the workflow - the addition of the substrate and the geometry relaxation - the presence of possibly spurious molecules can become problematic. In general, we filter out molecules from the predictions that are low-lying and either isolated or placed inside closed loops, as we have observed that they usually increase in height and form brighter, inconsistent features post-relaxation. An informed decision can be made after several trials, by observing how the molecule rearranges during the relaxations, and by visually evaluating its resulting features in the simulated images. 

Lastly, before bringing the surface into the picture, we also adjust eventual incomplete water molecules. Missing hydrogens are added so that the internal angles of their corresponding molecule are correct. Beyond this condition, the exact hydrogen positioning is not important at this stage, since the molecules will subsequently be rotated in searching for stable hydrogen bond frameworks. 

\subsection{Surface relaxation}

\begin{figure}[h]
	\centering
	\includegraphics[width=160mm]{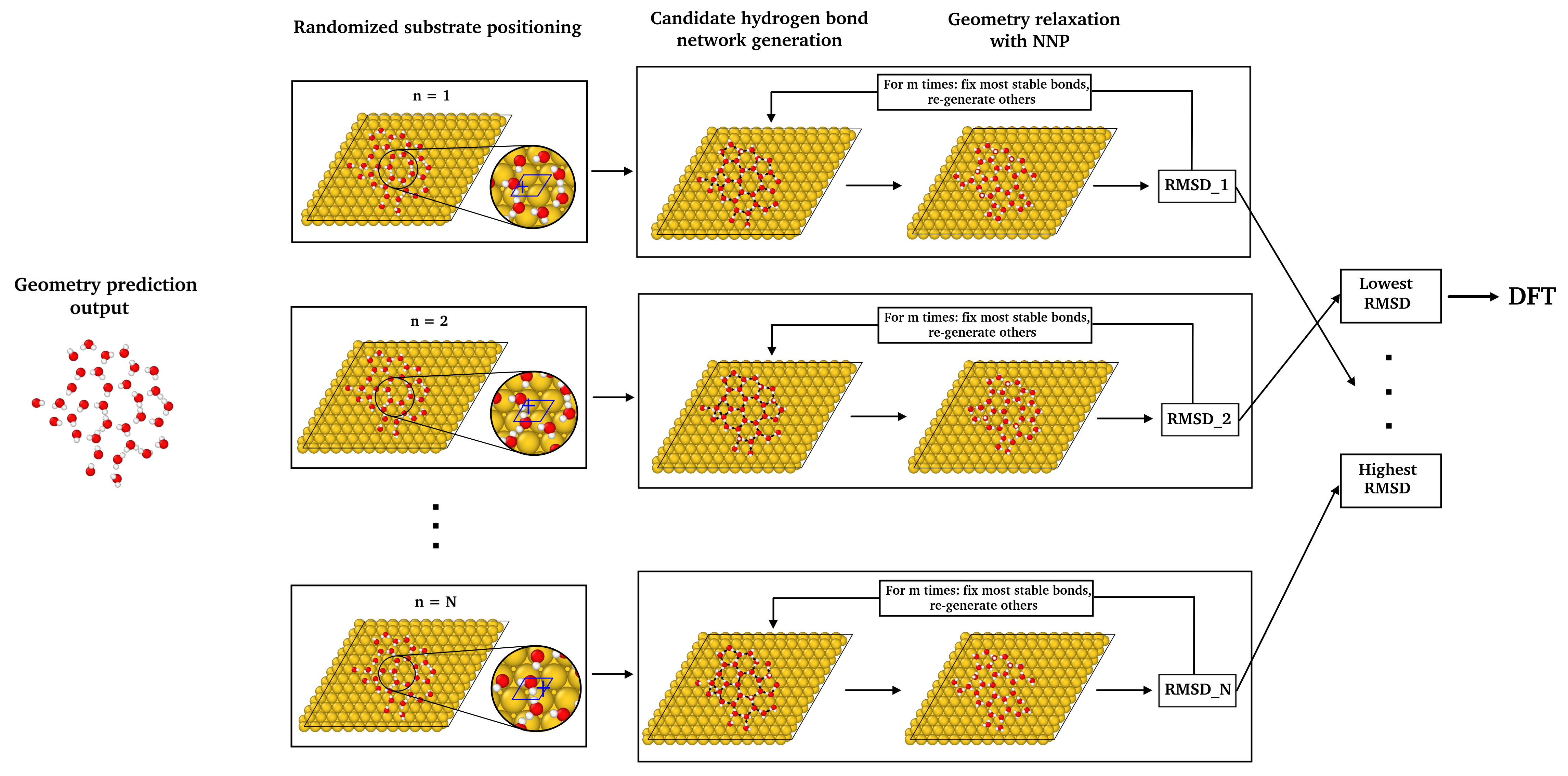}
	\caption{The guessing procedure to connect a ML geometry prediction to a DFT relaxation on a metal substrate. The geometry center of mass (shown as a blue cross in the snapshots) is positioned in a random position of the substrate unit cell, and the cluster is randomly rotated around z. In particular, the initial height of the cluster center of mass is taken randomly from an interval around the central peaks of the Figure S9 (a) and (b). For each of the starting positions, an iterative bond-creation / NNP-relaxation loop is then carried out. A tentative hydrogen bond network is constructed with the algorithm shown in \ref{fig:hydrogen_bonds}. The cluster is then relaxed using the NNP. The RMSD of the relaxed oxygens relatively to their initial position is then computed. At this point, more refinements iterations can be carried out, with the oxygens that did move less than a certain threshold being fixed, and the rest of the network being constructed again. The samples RMSDs are then sorted and the sample with the lowest one is relaxed with DFT. In the case of bilayer nanoclusters, the only change is the addition of an hexagonal ice layer between the substrate and the prediction, whose orientation is also randomly sampled.}
	\label{fig:iterative_relaxation}
\end{figure}

\begin{figure}[h]
	\centering
	\includegraphics[width=70mm]{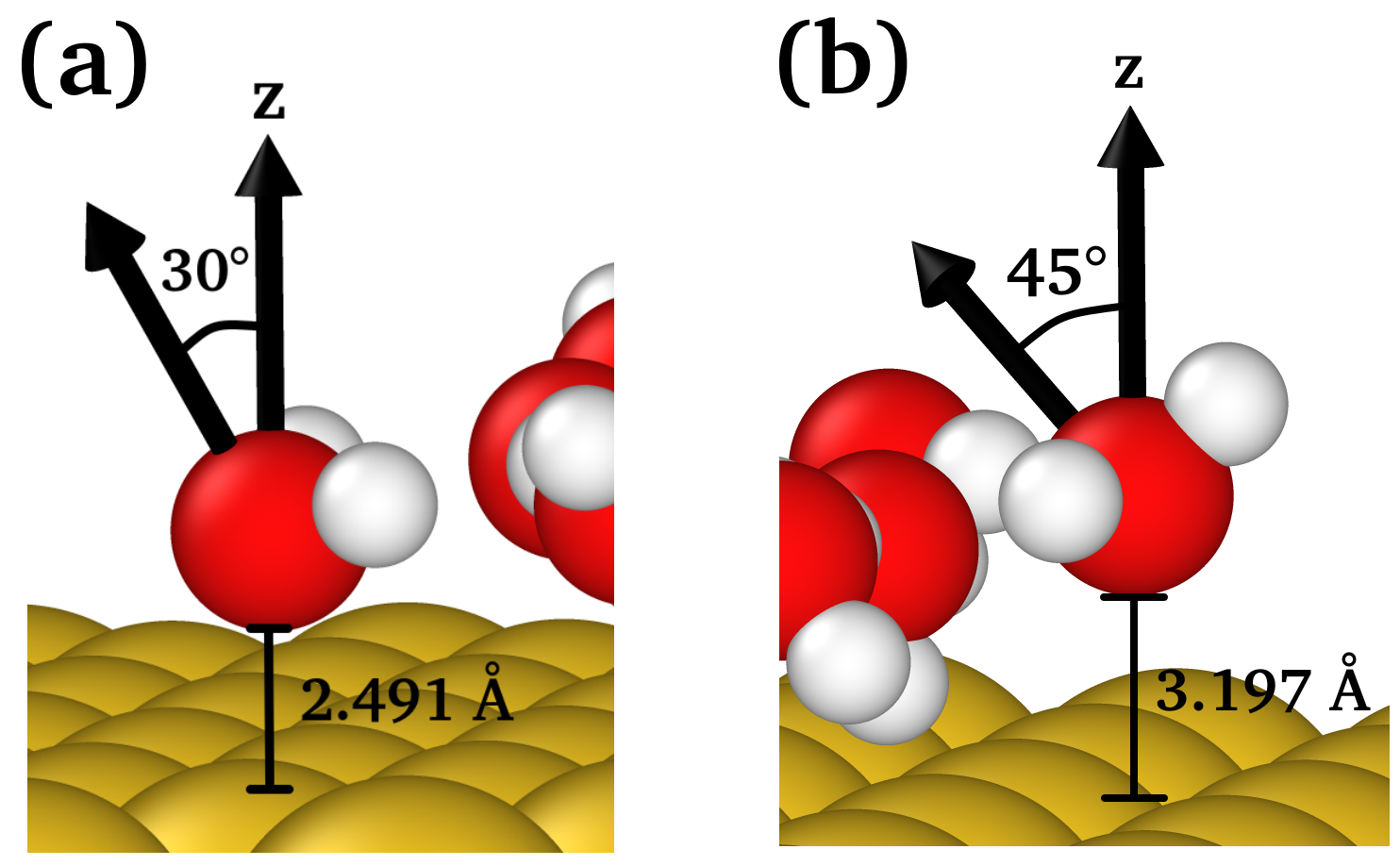}
	\caption{Two example molecular configurations from (a) the "below h1" zone of the Au(111) monolayer clusters histogram, and (b), the "h1-h2" zone of the same histogram.}
	\label{fig:au_angles_examples}
\end{figure}

In Fig.~\ref{fig:iterative_relaxation} in the SI we illustrate the process for searching for a stable combination of substrate positioning and hydrogen bond arrangement for a given monolayer cluster prediction. We follow the same procedure for bilayers, but also place below the predicted molecules an ice hexagonal layer as in the standard bilayer ice model \cite{Carrasco2012}, from which outer edges are removed to arrive at an isolated, non-periodic ice nanocluster. In the search, we consider the oxygens in the filtered ML prediction as the ground truth, and we then explore a large array of translations and rotations of a nanocluster center of mass on the surface. For each starting position, a different hydrogen bond network is then generated according to the algorithm in Fig.~\ref{fig:hydrogen_bonds}, also exemplified in Fig.~\ref{fig:hydrogen_bonds_example} in the SI. More specifically, hydrogen bonds are created starting from a randomly chosen water molecule, proceeding radially towards the others. A probability is assigned to each molecule to form one or two bonds with its neighbours according to its oxygen height above the surface, following simple statistical considerations from the ML datasets, as shown in Fig.~\ref{fig:water_distribution} and Fig.~\ref{fig:au_angles_examples} in the SI. Briefly, we found that in monolayers, low-lying molecules tend to be bonded with two neighbours, arranging in a horizontal position on the surface. Molecules at intermediate heights are instead more frequently positioned vertically (i.e. with one OH bond towards the surface), thus bonding with only one neighbour, while molecules with even higher oxygen $z$ coordinate are instead more diversely oriented. For bilayers, this latter situation applies to all the predicted molecules, which are effectively decoupled from the metal. In the bottom layer, waters are allowed to rotate upwards, if standing below a molecule that was positioned horizontally. Once the position and the bonds of each configuration have been set, we do a geometry optimization using the previously described NNPs. The NNP-relaxed structures are then ranked based on how much the oxygens were displaced compared to their ML prediction ground truth. The highest ranked structure is finally relaxed using DFT, and its Hartree potential is used for simulating the AFM images to be compared with experiment. 

\begin{figure}[h!] 
\begin{center}
\centering
\includegraphics[width=160mm]{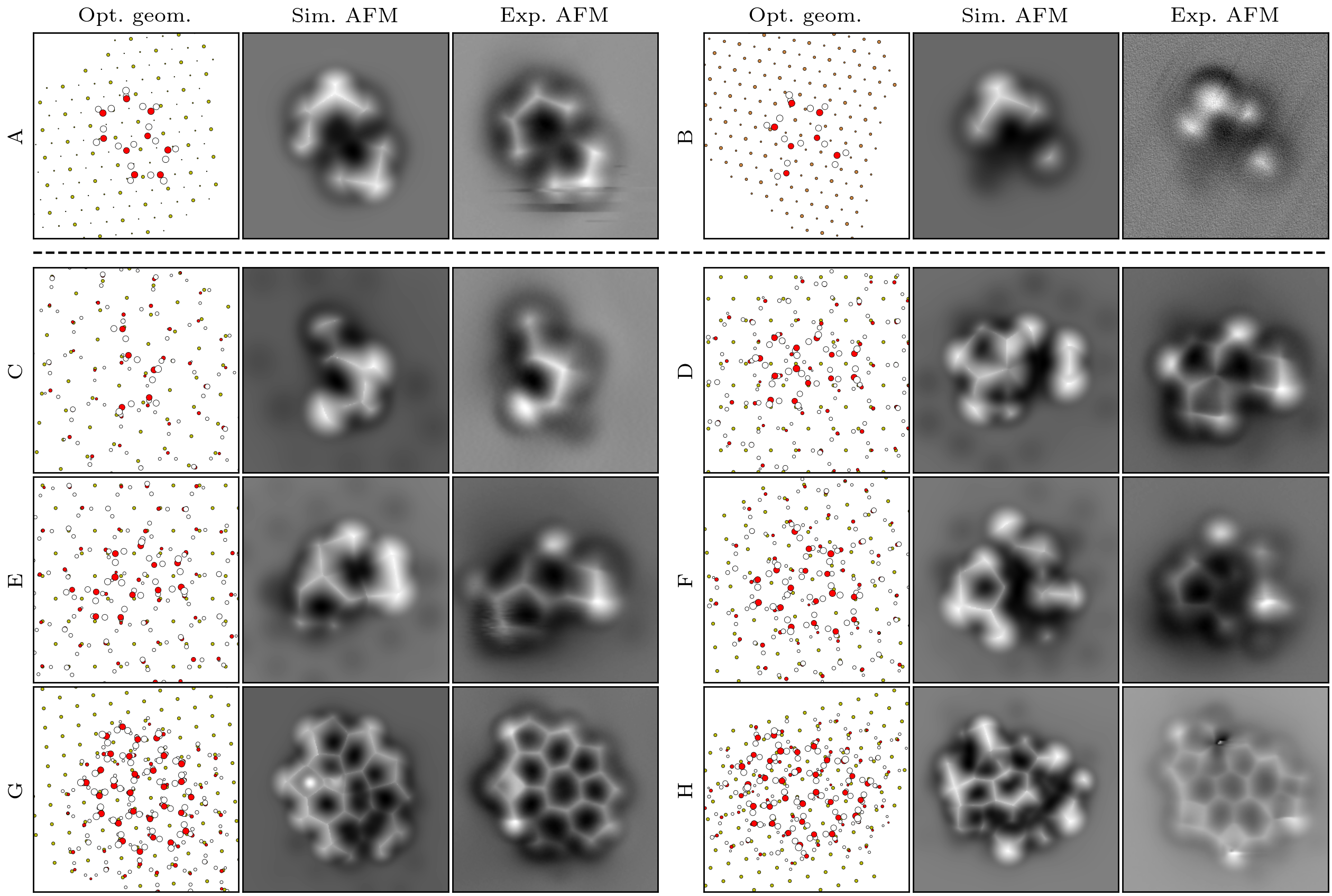}
\caption{On-surface geometries optimized with an NNP, corresponding simulated AFM images, and comparison to experiment. The rows A-H correspond to the experiments with same labels in Fig.~\ref{fig:exp_comp}. On each row, on the left is the final optimized geometry, in the middle is the simulated AFM image, and on the right the experimental image.}
\label{fig:nnp_sims}
\end{center}
\end{figure}

\begin{figure}[h]
	\centering
	\includegraphics[height=210mm]{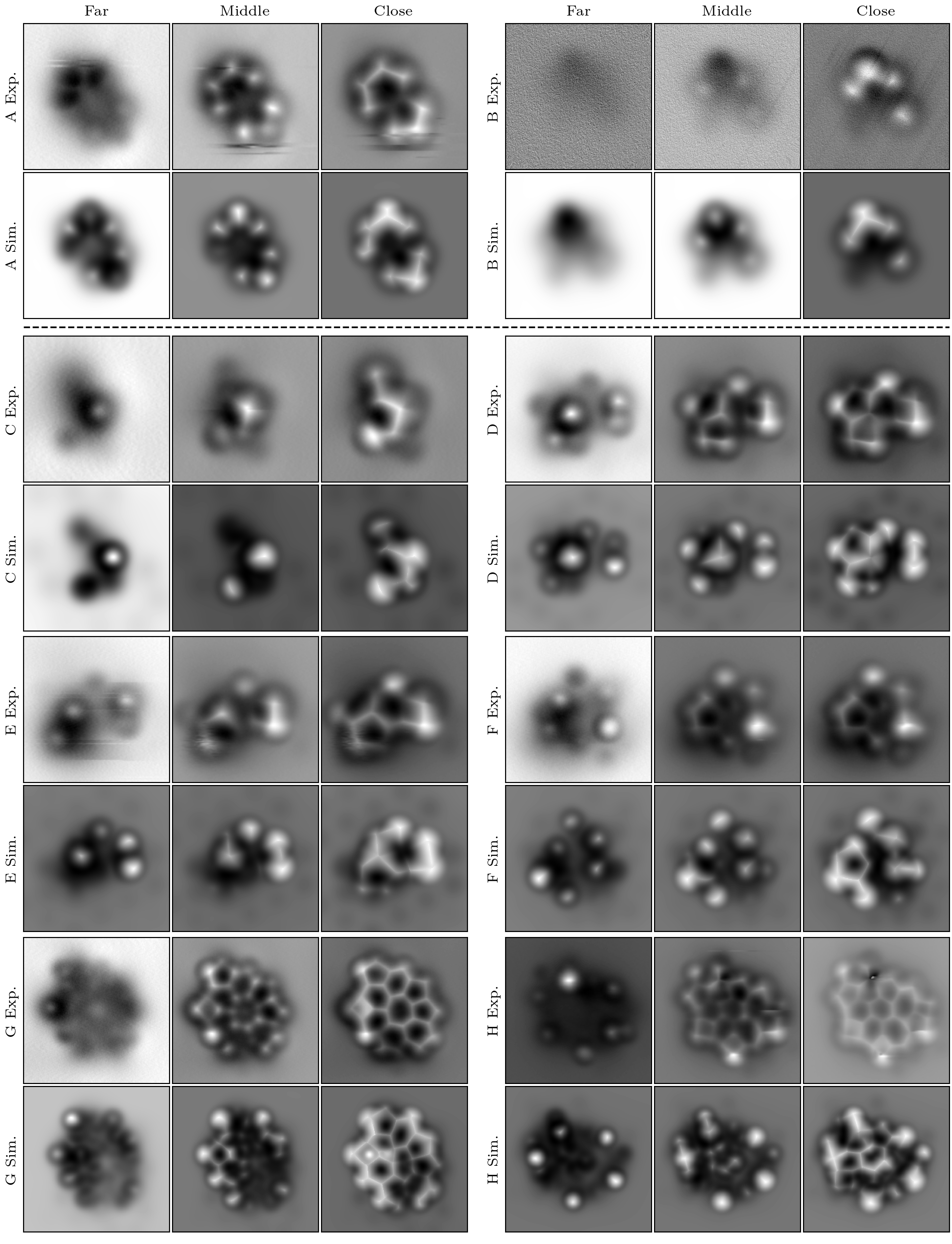}
	\caption{Extended comparison of the experimental images and simulations of the optimized geometries.}
	\label{fig:results_exp_opt}
\end{figure}

In Fig.~\ref{fig:nnp_sims}, we compare the surface-relaxed geometries, their AFM simulations and the experimental images. We do an extended comparison over three different tip-sample heights in Fig.~\ref{fig:results_exp_opt} in the SI. Generally, we find the agreement between simulations and experiments to be still very good after the addition of the underlying bottom ice layer and metal surface. In addition, we only observe minimal ionic movements during the DFT runs, confirming the accuracy of our NNPs. The agreement is particularly well-preserved for samples A, B and C, whose small size allowed the positioning and bonding search procedure to thoroughly explore most viable combinations. For the bigger clusters, we notice instead slight readjustments of the molecules, especially in their $z$ coordinate. This is to be expected, and it results from the higher complexity of these bigger structures, where even small errors in the oxygen positions or slight differences in the choice of hydrogen bonds could have global effects on the cluster geometry.

\section{Discussion}
We notice that structure predictions from samples A, E, F and G possess a similar final orientation of the underlying lattice, despite not having imposed such a rotational constraint on the clusters during the search. Certainly, in such amorphous clusters, multiple bonding configurations and center of mass rotations could still result in visually similar and stable geometries. Nevertheless, as the experiments were effectively conducted on an individual Au(111) clean surface, it is a promising result to retrieve the correct orientation. Sample B's final surface orientation was also found to be matching with the experimental one. This was the only sample measured on Cu(111), but STM images of its surface had been taken, and they allowed us to confirm the matching. 

\begin{figure}
	\centering
	\includegraphics[width=160mm]{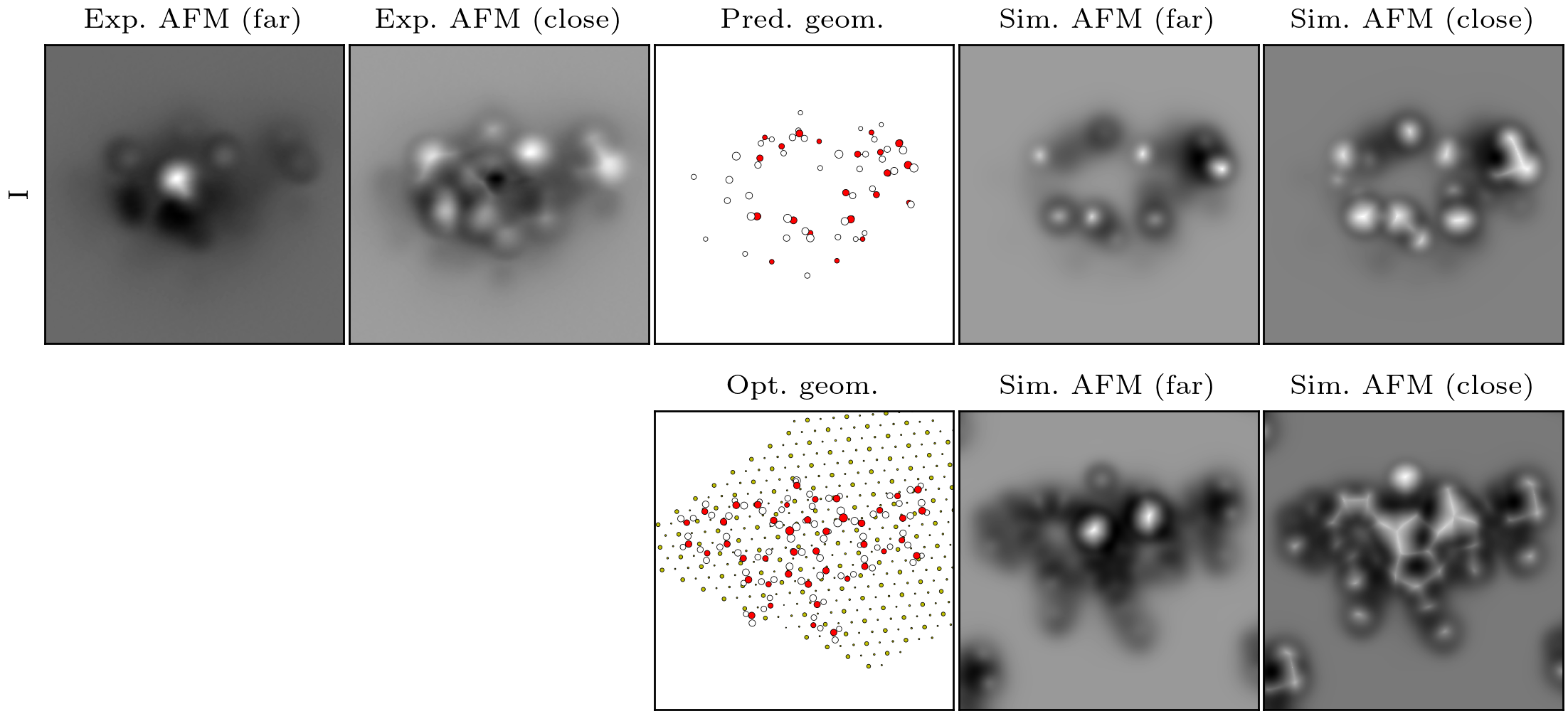}
	\caption{Additional experiment with a prediction and an optimized geometry, and the corresponding simulations.}
	\label{fig:results_extra}
\end{figure}

We also want to highlight an additional experiment, showed in Fig.~\ref{fig:results_extra} in the SI, for which the geometry prediction failed to extract the complete molecular structure, especially in the center of the cluster. To better understand what caused the ML models to struggle, we attempted to manually build a tentative geometry, which is also shown in the bottom row of the same figure. In all probability, the presence of very high and low molecules next to each other in a concentrated region of the cluster constituted the confounding factor for the ML predictions. The situation might have further been worsened by the lowest images being affected by tip-sample interactions, which may have altered the cluster geometry. In future improvements, we aim to address both problems. Firstly, we will ensure incorporation of similar configurations in our dataset, as a post-processing step on some of the randomly generated clusters. In general, as we analyze more experimental images and thus uncover more problematic patterns, we will correspondingly expand our dataset generation to include them. Secondly, tip-relaxation effects could be directly included in the creation of the datasets, providing the ML with educated guesses about structural changes of the clusters depending on the image height.

\section{Conclusions}

In conclusion, we have demonstrated the effectiveness of our workflow for structure discovery in three-dimensional ice nanoclusters and validated it by comparison to experimental measurements. In contrast to previous efforts in approaching similar problems, we now automate the structure prediction, the hydrogen bond arrangement and the substrate relaxation, leveraging recent advancements in machine learning which allow us to iterate through tentative structures much more efficiently compared to traditional DFT. Our work enables the partnering of high-resolution AFM with accurate simulations of a wide variety of, as yet, unexplored complex ice systems, which have generally been inaccessible due to the sheer difficulty in their interpretation. This will broaden and accelerate our understanding of heterogeneous ice nucleation and more generally of water-surface interactions, a fundamental concept in numerous scientific and technological domains.

\section{Methods}

\subsection{AFM experiments}
The experiments were performed on two separate non-contact AFM/STM systems (Createc, Germany) at 5 K using qPlus sensors equipped with a tungsten (W) tip (spring constant $k_0 \simeq \SI{1800}{N/m}$, resonances frequencies of $f_0 \simeq \SI{29.1}{kHz}$, and quality factor $Q \simeq 10^5$). All the AFM frequency shift ($\Delta f$) images were obtained with the CO-terminated tips in constant-height mode. The CO-tip was obtained by positioning the tip over a CO molecule on the Au(111) surface at a set point of \SI{100}{mV} and \SI{10}{pA}, followed by increasing the current to \SI{400}{pA}. The CO functionalization on Cu(111) was achieved by positioning the tip over a CO molecule at the set-point of \SI{8}{mV} and \SI{100}{pA}, followed by ramping the sample bias from zero to \SI{2.6}{V} with feedback off. The oscillation amplitude of experimental AFM imaging was \SI{50}{pm} (cluster B on Cu(111)) or \SI{100}{pm} (all experiments on Au(111)).  The Au(111) and Cu(111) single crystal were purchased from MaTeck. The Au(111) surface was cleaned by repeated Ar$^+$ ion sputtering at \SI{1}{keV} and annealing at about \SI{700}{K} for multiple cycles. The Cu(111) surface was prepared by repeated Ne$^+$ ion sputtering at \SI{750}{eV} followed by annealing at about \SI{820}{K}. The ultra-pure H$_2$O (Sigma Aldrich, deuterium-depleted, \SI{1}{ppm}) was used and further purified under vacuum by 3-5 freeze-and-pump cycles to remove remaining gas impurities. The water molecules were deposited in situ onto the surface held at \SI{5}{K} through a dosing tube and followed annealing at \SI{77}{K} for \SI{10}{min}.

\subsection{AFM simulations}

The AFM simulations are performed using the Probe Particle Model (PPM) \cite{hapala2014} code using a Lennard-Jones interaction and an electrostatic interaction calculated from the Hartree potentials of the samples obtained from the DFT calculations detailed below. The only exception to this is the simulations on the predicted geometries (Fig.~\ref{fig:exp_comp}), where only the Lennard-Jones is used, since the prediction only contains the geometry and atom types without any electrostatics information. The default built-in Lennard-Jones parameters in PPM are used.

In order to augment the dataset and make the trained machine learning models more robust, the simulation is performed multiple times for the same sample while varying some of the simulation parameters. Following the example of the QUAM-AFM dataset \cite{carracedo-cosme2022}, we vary the lateral spring constant $k_\mathrm{xy}$ and the oscillation amplitude $A$. In addition, we vary the closest tip-sample distance $d_\mathrm{ts}$ and the lateral equilibrium position of the probe particle $t_\mathrm{xy}$. In the QUAM-AFM dataset the parameters take all combinations of predetermined values, in total 24 different combinations for the two parameters. In our case where we have 4 parameters to vary, the number of different combinations would grow very large, so we instead choose to do a fixed number of 10 simulations for each sample, randomly picking the parameters in set ranges from a uniform distribution. The set ranges are $0.2 - 0.5\ \si{N/m}$ for $k_\mathrm{xy}$, $0.4 - 2.0\ \si{\angstrom}$ for $A$, $t_\mathrm{xy}$ is in a disk of radius $\SI{0.3}{\angstrom}$, and $\Delta d_\mathrm{ts} = \SI{0.5}{\angstrom}$. The average of $d_\mathrm{ts}$ is chosen by eye such that sharp features like ones seen in real AFM images at close approach are seen in the simulated images. The simulations are performed at 15 tip-sample distances at $\SI{0.1}{\angstrom}$ step producing 3D stacks of AFM images.

\subsection{Geometry prediction model}

The atomic geometry prediction model follows closely our previous work on reconstructing molecule graphs from AFM images \cite{Oinonen2022} with some modifications to make the model more general. The basic structure of the model stays the same: there is a convolutional neural network (CNN) that first predicts the positions of the atoms from the AFM image stack, and a graph neural network (GNN) that uses the predicted positions along with the AFM images to construct a labelled molecule graph. The biggest differences to the original model are that the CNN and GNN networks are now completely separate without any shared layers between them, the CNN is modified to allow arbitrary size inputs in the $z$-dimension, and the GNN is simplified to label and connect the molecule graph in one shot for the whole graph instead of iteratively for each atom. These modifications are explained in more detail in the following.

\subsubsection{Atom position prediction}

The atom positions are predicted using an Attention U-Net CNN\cite{schlemper_2019, ronneberger_u-net_2015} modified to accept variable size inputs while producing fixed-size outputs in $z$. In the U-Net architecture, the input image is first passed through an encoder that has a series of CNN blocks interleaved with pooling layers that gradually down-sample the feature maps to a smaller size and then a decoder that gradually up-samples the feature maps back to the original size. Additionally, there are skip-connections between the corresponding stages of the encoder and decoder that allows information to propagate in the network at multiple scales and allows more efficient back-propagation of gradients. The Attention U-Net\cite{schlemper_2019} adds to the skip connections \emph{attention gates} that produce a map of coefficients in $[0, 1]$ for every pixel in a given feature map which is multiplied by that map of coefficients, therefore forcing the model to highlight the relevant regions in the feature map. The attention mechanism is useful by itself for improving the model performance, but they can also be used for gaining insight into what regions the model focuses on for making the prediction\cite{ranawat2021}, although here do not make use of this aspect of the model.

We construct here a variant of the attention-gate layer which modifies the size of the feature map in $z$ into a fixed size. Suppose we have a feature map $X$ in the middle of the network with size $K$ in the $z$-dimension, then we produce a new feature map $X'$ with size $K'$ by applying the operation
\begin{equation} \label{eq:attention}
    X'_{k'} = \sum_{k=1}^{K} \sigma(f_{k'}(X))_k \odot X_k \quad \forall k' \in \{1...K'\},
\end{equation}
where $f_{k'}$ is a convolution block with unique weights for each z-layer $k'$ in the output and with padding to retain the $z$-size throughout, $\sigma: \mathbb{R} \rightarrow [0, 1]$ is an activation function, and $\odot$ denotes element-wise multiplication. Here we choose to use 3 layers and zero-padding in the convolution block, and we use the sigmoid activation function $\sigma(z) = 1/(1 + \exp(-z))$. We add this layer to the output of the encoder and all of the skip connections, so that the decoder can work with fixed $z$-size feature maps at all scales. The value for $K'$ is a hyperparameter for which we choose here the values 3 for the encoder output and 3, 5, and 10 for the skip-connection outputs from smallest to largest scale. The encoder does not use any pooling in the z-dimension, so that the input AFM image stack can even have just a single $z$-layer.

\subsubsection{Graph construction}

The original graph construction model \cite{Oinonen2022} works in an iterative way, taking one of the predicted atom positions at a time and adding a corresponding node to the graph with associated edges corresponding to chemical bonds. The node labelling process is informed by a channel of information coming from a CNN that is shared with the U-Net. The structure of the model creates some problems for the predictions. First, the iterative nature of the model makes the predictions dependent on the order in which the atoms are added to the graph. Second, the shared weights between the two networks makes the training process more difficult because the two prediction tasks need to be balanced at the same time. Additionally, the way that the GNN uses absolute coordinates of the atoms makes the predictions dependent on the chosen coordinate system and the exact size of the AFM images, and especially makes the predictions close to the edge of the image less reliable \cite{Oinonen2022}.

\begin{figure}[ht!]
	\centering
	\begin{tikzpicture}[scale=1.2]
		\node[anchor=north west, inner sep=0, outer sep=0] at ( 0.0, 0.0) {\includegraphics[scale=1.2]{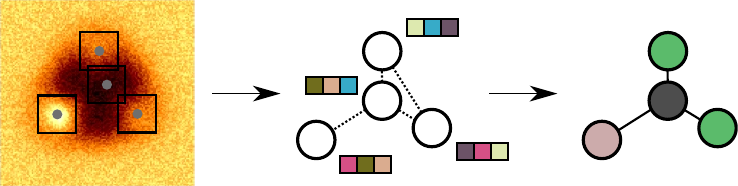}};
		\node[anchor=mid, inner sep=0, outer sep=0]        at ( 1.65, 0.25) {Select patches};
		\node[anchor=mid, inner sep=0, outer sep=0]        at ( 6.30, 0.25) {Initial graph};
		\node[anchor=mid, inner sep=0, outer sep=0]        at (11.10, 0.25) {Final graph};
		\node[anchor=mid, inner sep=0, outer sep=0]        at ( 4.10,-1.20) {CNN};
		\node[anchor=mid, inner sep=0, outer sep=0]        at ( 8.80,-1.20) {GNN};
	\end{tikzpicture}
	\caption{Schematic of the one-shot graph construction model. The process starts on the left with overlaying the found atom positions (gray dots) onto the AFM image and selecting rectangular patches around these positions. A CNN then turns the image patches into the initial embedding vectors (coloured squares) for every node of the graph, and edges (dashed lines) are added to graph based on the proximity of the nodes. Finally, a GNN processes the information in the node vectors and does a classification of the nodes into atom types and a binary classification on each edge whether it corresponds to a chemical bond.}
	\label{fig:one-shot-graph}
\end{figure}

Hence, we modify the graph construction network here in a way that addresses all of these issues (see schematic in Fig.~\ref{fig:one-shot-graph} in the SI). Instead of processing the whole stack of AFM images as a whole, we instead choose small patches from the AFM images for each of the predicted atoms based on the proximity of the atom coordinates to the coordinates of the AFM image pixels. The patches are processed with a CNN to produce fixed-size feature vectors for each of the atoms, and these feature vectors are then used as initial hidden vectors for the nodes in a GNN. The GNN mixes the information between the nodes for multiple rounds along edge connections based proximity between the atoms, and finally the nodes are classified by a multi-layer perceptron (MLP). This process simplifies the model by getting rid of the iteration for the nodes and makes it one-shot instead, with the AFM features gathered locally in a way that makes the predictions independent of the lateral size of the AFM image. Additionally, only relative coordinates are used within the network, so that the choice of the origin of the coordinate system is arbitrary.

Starting with the AFM images, let $\{r^q\}_{q=1}^N$ be the set of coordinates for the atoms produced by the U-Net model, and let $\{R_{ij}\}$ be the set of $xy$-coordinates for the voxels of the AFM image stack. For each of the atom coordinates $r_q$, we gather from the stack of AFM images a square patch of voxels whose $xy$-coordinates are within a cutoff distance $d_\mathrm{a}$: $\lVert R_{ij} - r_{ij}^q \rVert_{\infty} \leq d_\mathrm{a}$. We use the value $d_\mathrm{a} = \SI{1.125}{\angstrom}$ for the cutoff, which with a pixel resolution of $\SI{0.125}{\angstrom}$ creates patches of size $19 \times 19$. In the case where the atom is close to the edge of the image, the image is padded with zeros so that the patches have a constant size. This produces a set of smaller AFM images which are processed by a CNN to produce a fixed size feature vector for each atom. The CNN has three ResNet \cite{He2016} blocks with $2 \times 2$ pooling after each block, an attention gate layer similar to Eq.~\eqref{eq:attention} to reduce the 3D feature map down to just a single voxel, and a one final fully-connected layer that transforms the feature vector size to the one used inside the GNN. The three ResNet blocks have 12, 24, and 48 channels, respectively, and all have 2 layers.

The final feature vectors from the preceding CNN are used as the initial hidden vectors $h_{v}^0$ for each node labelled with $v \in \{1 \dots N\}$. The hidden vectors are updated for $n_\mathrm{t}$ iterations by a message-passing GNN \cite{gilmer2017}:
\begin{align}
    m_{vu}^t &= f_\mathrm{m}(h_v^{t-1}, h_u^{t-1}, r^u - r^v) \quad \forall u \in \mathcal{N}(v),\ \forall v \in \{1 \dots N\}, \label{eq:message} \\
    h_v^t &= f_\mathrm{h}\left( h_v^{t-1}, \sum_{u \in \mathcal{N}(v)} m_{vu}^t \right) \quad \forall v \in \{1 \dots N\},
\end{align}
where $t \in \{1 \dots n_\mathrm{t}\}$, $f_\mathrm{m}$ is an MLP, $f_\mathrm{h}$ is a gated recurrent unit (GRU) \cite{cho2014}, and $\mathcal{N}(v)$ denotes the set of neighbours for the node $v$. Note that the message function in Eq.~\eqref{eq:message} uses the relative coordinates between the nodes, which makes the update iteration translationally invariant. The set of neighbours of a node is decided based on proximity to other nodes: given a cutoff distance $d_\mathrm{e}$, the set of neighbours for a node $v$ is $\mathcal{N}(v) = \{u\ |\ \lVert r^u - r^v \rVert_2 \leq d_\mathrm{e} \}$. Here we use the value $d_\mathrm{e} = \SI{3}{\angstrom}$ for the cutoff, which is enough to capture all possible bonding distances between the atoms, including hydrogen bonding. Additionally, we choose $n_\mathrm{t} = 5$, $|h_v^t| = |m_{vu}^t| = 40$, and $f_\mathrm{m}$ has two hidden layers of size $196$.

The final classification of the node types is done by another MLP $f_\mathrm{c}$:
\begin{equation}
    c_v = f_\mathrm{c}(h_v^{n_\mathrm{t}}) \quad \forall v \in \{1 \dots N\}.
\end{equation}
The final layer of $f_\mathrm{c}$ is followed by a softmax activation so that $y_v$ is a probability distribution over the node classes. The loss for the classification task is the cross entropy loss, $L(c_v, c'_v) = - \sum_{i=1}^C c'_{v,i} \log c_{v,i}$, where $C$ is the number of classes, and $c'_v$ is a one-hot vector for the ground-truth class of node $v$. We choose to use one hidden layer of size $196$ in $f_\mathrm{c}$.

In addition to labelling the nodes, the edge connections between the nodes, corresponding to the bonds between the atoms, can be constructed. The bonds are not used here in practice for the subsequent simulations, but the method is described here for generality. The basic idea is to take the neighbour connections $\mathcal{N}(v)$ between the nodes and do a binary classification for each one on whether it corresponds to an edge in the final graph or not. To this end, in addition to maintaining a hidden vector for each node, we also maintain a hidden vector $g_{uv}^t$ for each (unordered) pair of neighbouring nodes $(u,v) \in E = \{(u, v)\ |\ u \in \mathcal{N}(v)\}$. The hidden vector is initialized to the average of the node hidden vectors, $g_{uv}^0 = (h_u^0 + h_v^0) / 2$, and then updated on each iteration of the GNN as
\begin{equation}
    g_{uv}^t = f_\mathrm{g}(g_{uv}^{t-1}, (m_{uv}^t + m_{vu}^t) / 2) \quad \forall (u, v) \in E,
\end{equation}
where $t \in \{1 \dots n_\mathrm{t}\}$, and $f_\mathrm{g}$ is a GRU. The final classification of the edge connections is done as
\begin{equation}
    e_{uv} = f_\mathrm{e}(g_{uv}^{n_\mathrm{t}}) \quad \forall (u, v) \in E,
\end{equation}
where $f_\mathrm{e}$ is an MLP with sigmoid activation in the final layer. The loss for the prediction is the binary cross-entropy loss. Like with the node classifier, we use one hidden layer of size $196$ in $f_\mathrm{e}$. The activation function for all of the layers in the model is the ReLU function.

\subsubsection{Model training}

During training, the AFM images are preprocessed in several ways following previous work \cite{Oinonen2021}, including normalization, random noise, cutouts, pixel shifts, random background gradients, and random rotations and reflections. Here we also randomize the $z$-size of the AFM image stack between 1 and 15 slices and randomize the starting slice between 1 and 5. The simulations additionally have 10 different random parameter sets for each sample, as described above. The samples are divided into several shards, and for each training epoch, one of the parameter sets is chosen at random for each shard.

The graph construction model is trained separately from the position prediction model. In order to account for the fact that the predicted positions have some uncertainty in them, we add Gaussian noise ($\sigma = \SI{0.08}{\angstrom}$) to the input node positions during the training of the graph construction model. The model parameters are optimized with the Adam optimizer\cite{kingma2014}, using the default momentum parameters. The models are trained for 1000-1500 epochs until the loss does not improve significantly anymore. The final model parameters are chosen from the epoch with the lowest validation loss.

It may be of interest that we also tried first training the models on a larger dataset with more elements \cite{Oinonen2021} and then fine-tuning on the water-only dataset. However, we found in practice that these models did no better or worse than ones trained from the beginning on the water dataset, as measured by the training and validation losses.

\subsection{DFT calculations}

DFT calculations were conducted with the Vienna Ab-initio Simulation Package (VASP) \cite{Kresse1996a, Kresse1996b}, modeling core electrons with projector augmented wave (PAW) potentials and expanding valence electrons with plane-waves with Ecutoff = 500 eV. The non-local van der Waals-density functional optB86b-vdW-DF \cite{Klime2011, RomnPrez2009, Dion2004} was utilized, as it has been shown to accurately describe the adsorption of water molecules on metal substrates \cite{Forster2011, Lew2011, Guo2014, Liriano2017, Dong2018}.

Depending on the size of the water cluster in consideration, a 9 $\times$ 9, 11 $\times$ 11 or 13 $\times$ 13 Cu/Au(111) supercell was selected, using 3 atomic layers and a vacuum separation of 20 Å along the slab perpendicular direction. Convergence tests of the k-grid showed that $\Gamma$ point calculations were already accurate at the meV level for the considered supercells.  

\subsection{Neural network potentials}

NNPs were employed via NequIP \cite{Batzner2022}, which allows building of an E(3)-equivariant NNP from a reference dataset of \emph{ab-initio} calculations. By leveraging euclidean neural networks from e3nn \cite{e3nn_paper} and by utilizing both scalar and higher-order tensor atomic features, the NequIP architecture reaches state of the art accuracy and data efficiency. 

\begin{figure}[h]
	\centering
	\includegraphics[width=160mm]{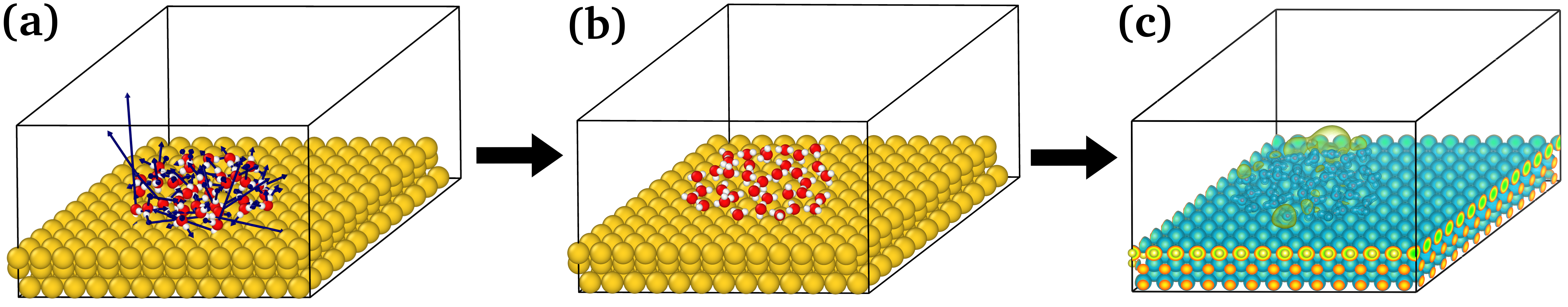}
	\caption{Construction of the dataset for training the ML geometry prediction model. In (a), between 5 and 50 water molecules are randomly positioned on the substrate, with varying starting heights. In bilayers, an hexagonal ice layer is also added below the molecules. In (b), we do a geometry relaxation using the NNP. In (c), we run single-point DFT on the NNP-relaxed structure to extract the electrostatic potential and the charge density distribution, which will be the input for training the ML geometry prediction model. The procedure was carried out for 2000 monolayers and 2000 bilayers for Au, and for 1850 monolayers for Cu.}
	\label{fig:dataset_ml}
\end{figure}

\begin{figure}[h]
	\centering
	\includegraphics[width=160mm]{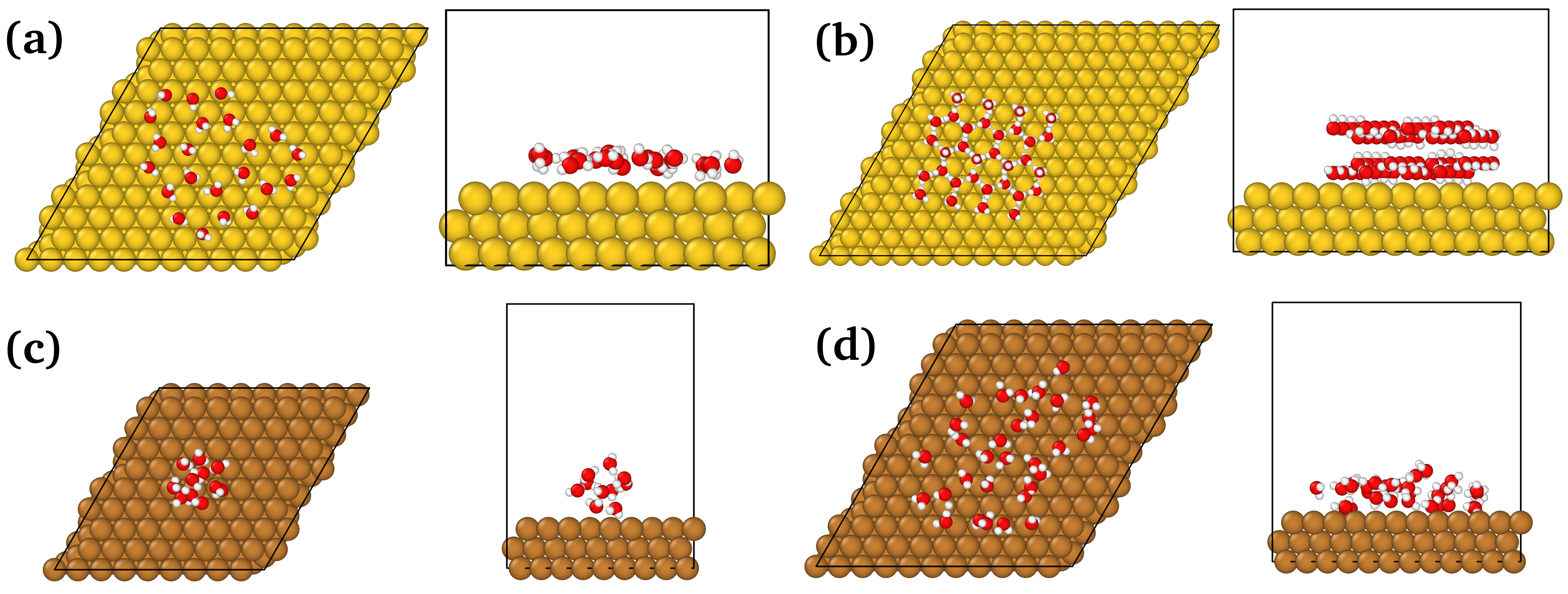}
	\caption{Example input structures from the 700 single-point DFT calculations used for training the Au(111) and Cu(111) NNPs. The same water structures were used on both metal surfaces. In (a) and (d), preliminary predictions from a previous machine learning model were distorted into highly strained configurations, to make the NNP robust. In particular, both hydrogen and covalent bonds were either stretched or compressed, molecules were randomly rotated and random translations were applied at both the atomic and molecular level. To further increase the diversity of the structures, we then added structures from the available ice types in GenIce \cite{Matsumoto:2017bk} as shown in (b). For the same reason, in (c) we show a water cluster taken from the Water Cluster Database \cite{Rakshit2019}. For randomly selected cases in (b) and (c), we also applied the same distortions discussed for (a) and (d).}
	\label{fig:dataset_nnp}
\end{figure}

\begin{figure}[h]
	\centering
	\includegraphics[width=160mm]{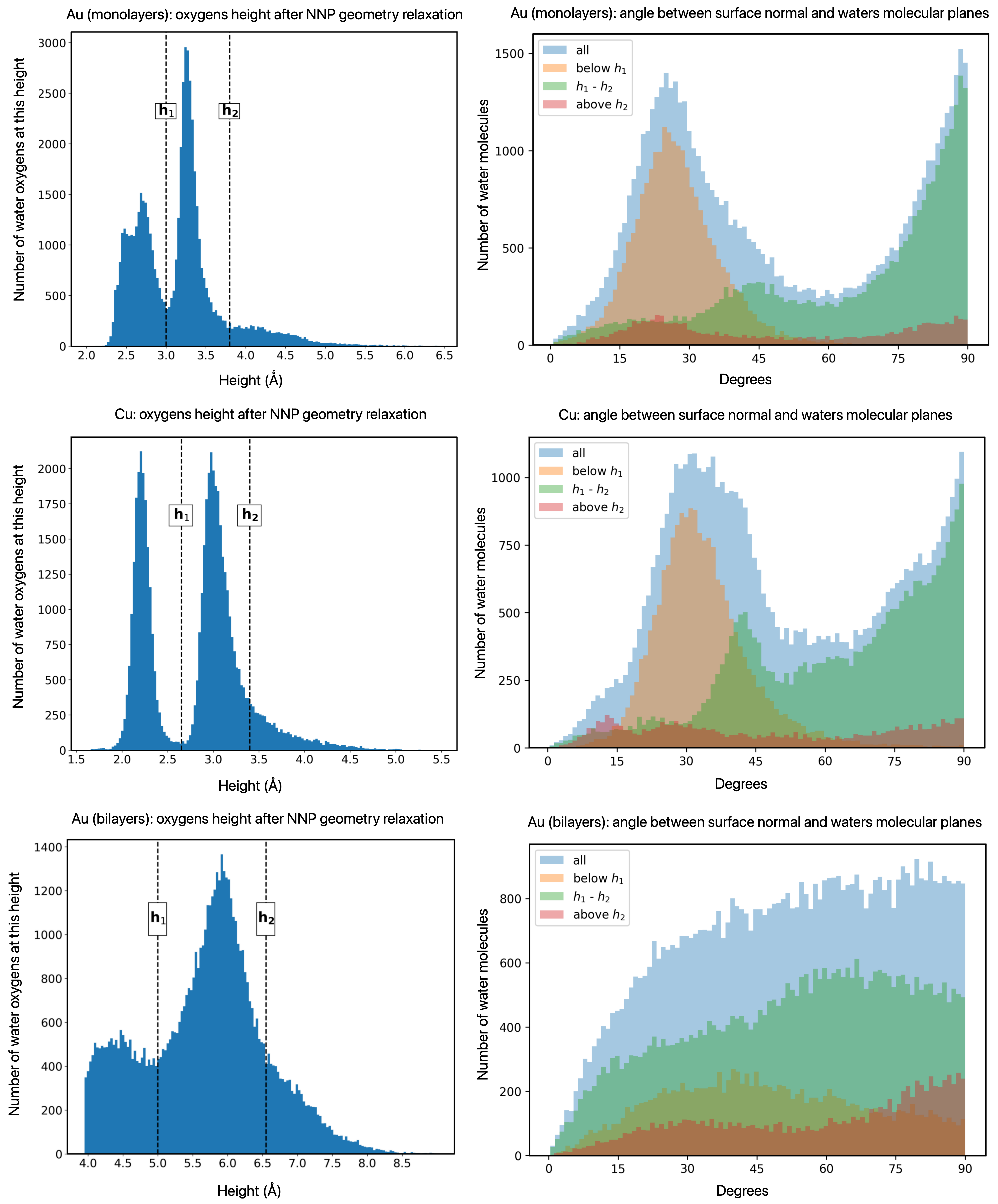}
	\caption{Analysis of the NNP-relaxed geometries for the 2000 monolayer clusters on Au(111) (top), the 1850 monolayer ones on Cu(111) (middle), and the 2000 bilayer ones on Au(111) (bottom), generated as shown in Figure \ref{fig:dataset_ml}. In the left column, we report the height distribution above the surface. In the right column, the distribution of the angle between the water molecular planes and the substrate normal vector for the molecules in various height bins.}
	\label{fig:water_distribution}
\end{figure}

Initially, two NequIP models were trained for Au(111) and Cu(111) on 700 structures exemplified in Fig.~\ref{fig:dataset_nnp}, which provided a diverse variety of both high and low energy structures to avoid overfitting and making a robust interatomic potential. These preliminary NNPs were then used to geometry relax a series of randomized water clusters to build the ML dataset, as shown in Fig.~\ref{fig:dataset_ml}. 

The relaxations were performed with ASE \cite{ase-paper} BFGS minimizer, using a force tolerance of 3 meV/ {\AA}. After carrying out single-point DFT runs on the relaxed clusters, the NNPs were finally retrained on both their initial dataset and on a portion of these new structures, which were more representative of our target distribution compared to the initial dataset. For both the Au(111) and the Cu(111), we utilized 1850 data points, a rotation order of $l=2$, a batch size of 2, a learning rate of 0.0075 and a \textit{PerSpeciesL1Loss} for the force loss term. For the Au(111) model, we obtained validation energy and force Mean Absolute Errors (MAE) of respectively 0.414 meV / atom and 0.00558 meV/ {\AA} (of which \textit{H\_f\_mae} = 0.0147, \textit{O\_f\_mae} = 0.0184 and \textit{Au\_f\_mae} = 0.00362). For the Cu(111) model, the final MAEs were instead 0.859 meV / atom for the validation energy and 0.00950 meV/ {\AA} for the validation force (of which \textit{H\_f\_mae} = 0.0160, \textit{O\_f\_mae} = 0.0215, and \textit{Cu\_f\_mae} = 0.00789).

\section{Data and materials availability}
All relevant data supporting the findings of this study are available in the Supplementary
Information. The source code and training data for the machine learning models is available under \url{https://github.com/SINGROUP/ml-spm}. 

\section {Conflict of interest}
The authors declare no competing financial interest.

\section{Author Contributions}
F.P. and N.O. created the data, performed the machine learning and atomistic simulations, and wrote the first version of the manuscript. Y.T., D.G., C.X. and S.C performed the experiments. All authors reviewed and commented on the manuscript. A.S.F. supervised the project.

\section {Acknowledgements}
The authors would like to thank Darina Andriychenko for her contribution to data collection, and Fedor Urtev for his assistance in the ML analysis. This work was supported by World Premier International Research Center Initiative (WPI), MEXT, Japan and by the Academy of Finland (Projects No. 347319, 347611, 346824). The authors acknowledge the computational resources provided by the Aalto Science-IT project and CSC, Helsinki.

\newpage

\section{Supplementary Information: Structure discovery in Atomic Force Microscopy imaging of ice}

\def\capt{Full sets of experimental AFM images. Images with red outline were produced by interpolation. The blue scale bar has a length of \SI{5}{\angstrom}.}
\begin{figure*}[hp!]
\centering
\includegraphics[width=150mm]{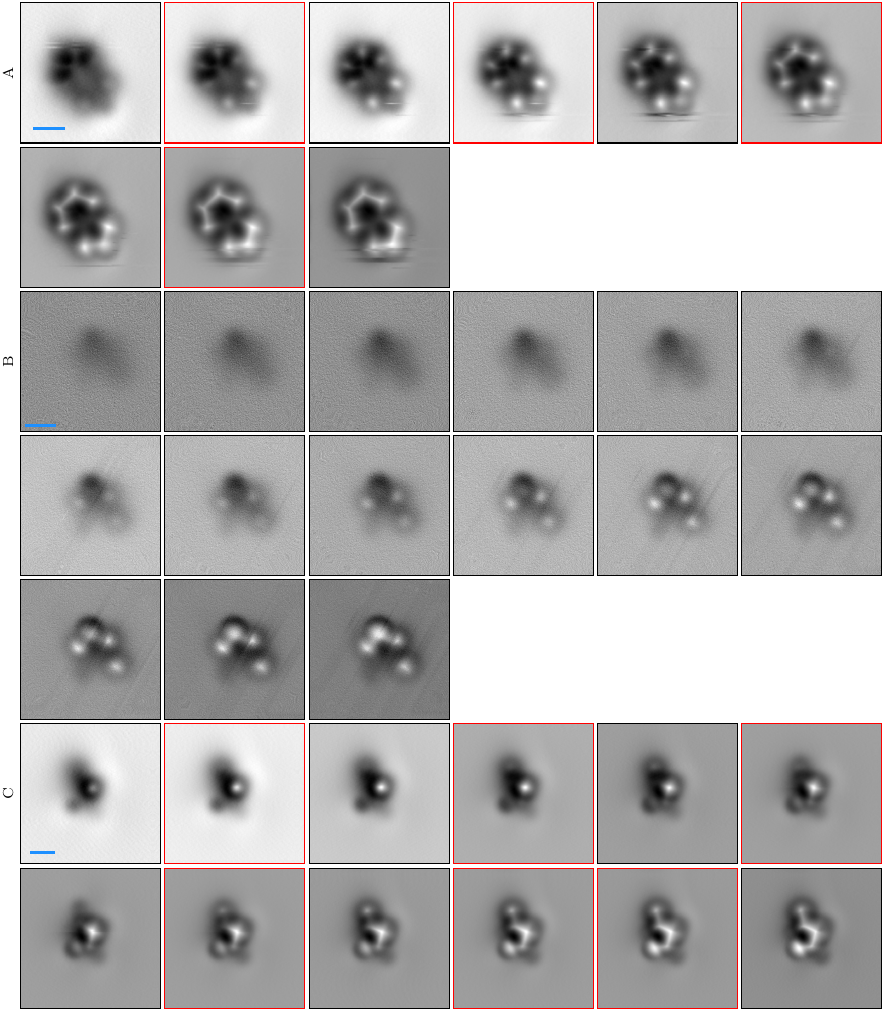}
\caption{\capt}
\label{fig:afm_full}
\end{figure*}

\begin{figure*}[hp!]
\ContinuedFloat
\centering
\includegraphics[width=150mm]{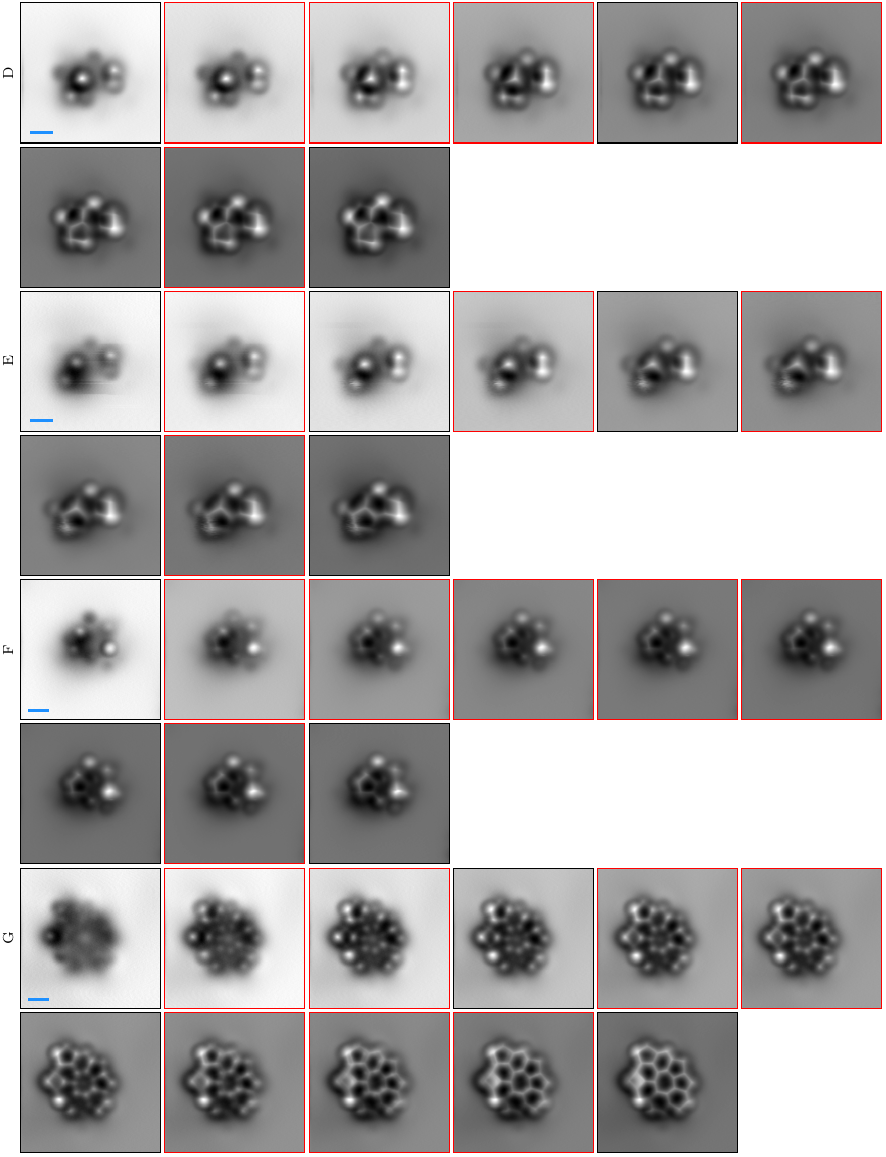}
\caption{(Cont.) \capt}
\label{fig:afm_full2}
\end{figure*}

\begin{figure*}[ht!]
\ContinuedFloat
\centering
\includegraphics[width=150mm]{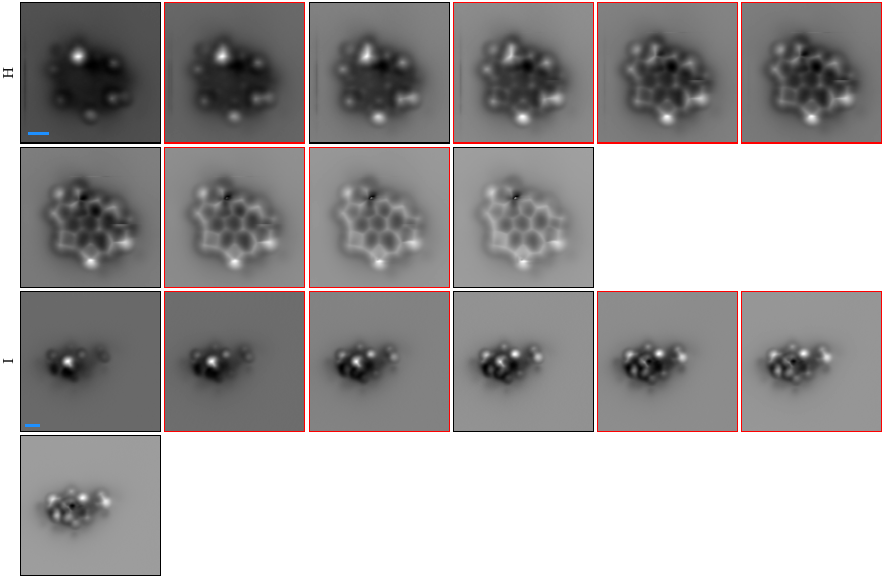}
\caption{(Cont.) \capt}
\label{fig:afm_full3}
\end{figure*}

\clearpage

\subsection{Effect of the AFM image stack size}
\label{sec:results_stack_size}

\begin{figure}
	\centering
	\includegraphics[width=160mm]{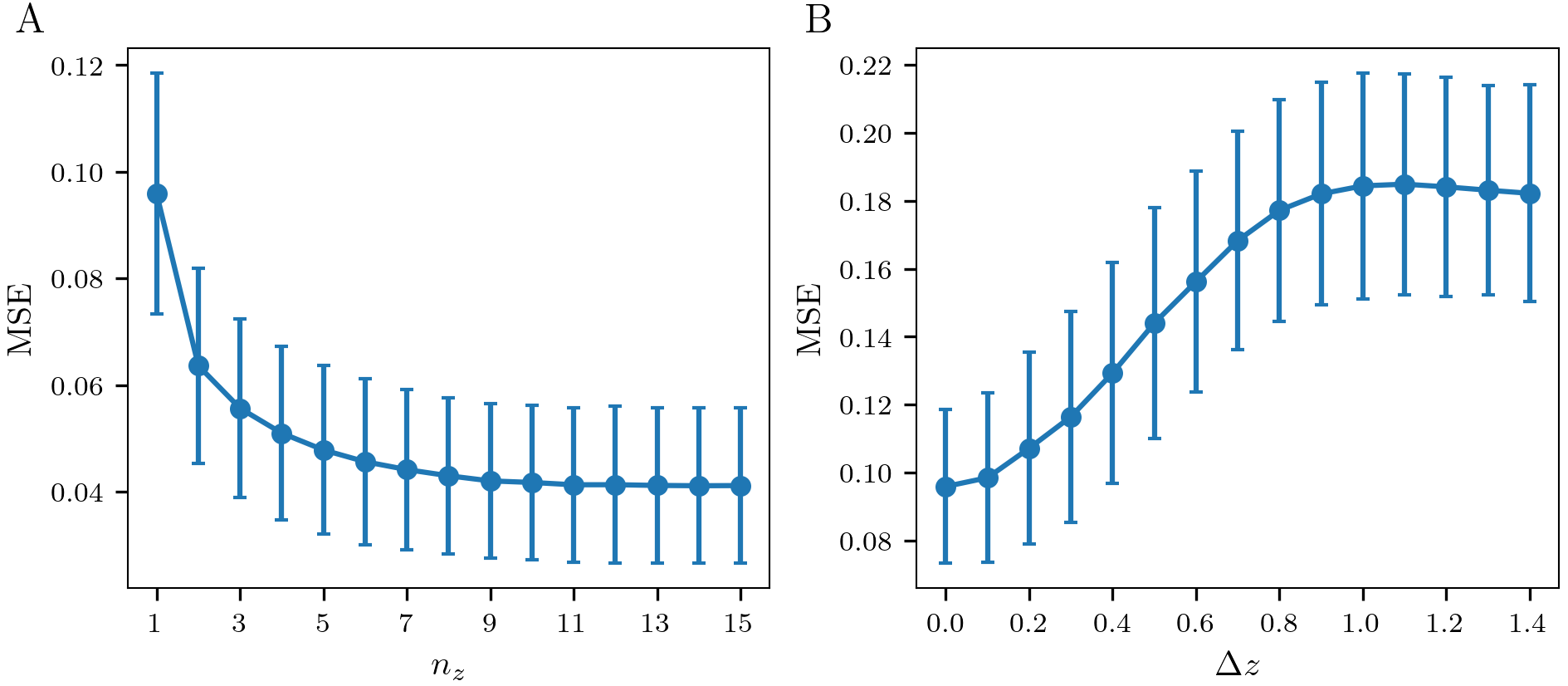}
	\caption{Position prediction mean squared error loss on the test set (A) as a function of the stack size $n_z$ and (B) for a single-image stack as a function of the distance offset $\Delta z$. The lengths of the vertical bars correspond to one sample standard deviation to both directions. It should be noted that the distance $\Delta z = 0$ does not correspond to any single tip-sample distance, but rather is the average closest distance in the sample distribution.}
	\label{fig:loss_z}
\end{figure}

\begin{figure}
	\centering
	\includegraphics[width=125mm]{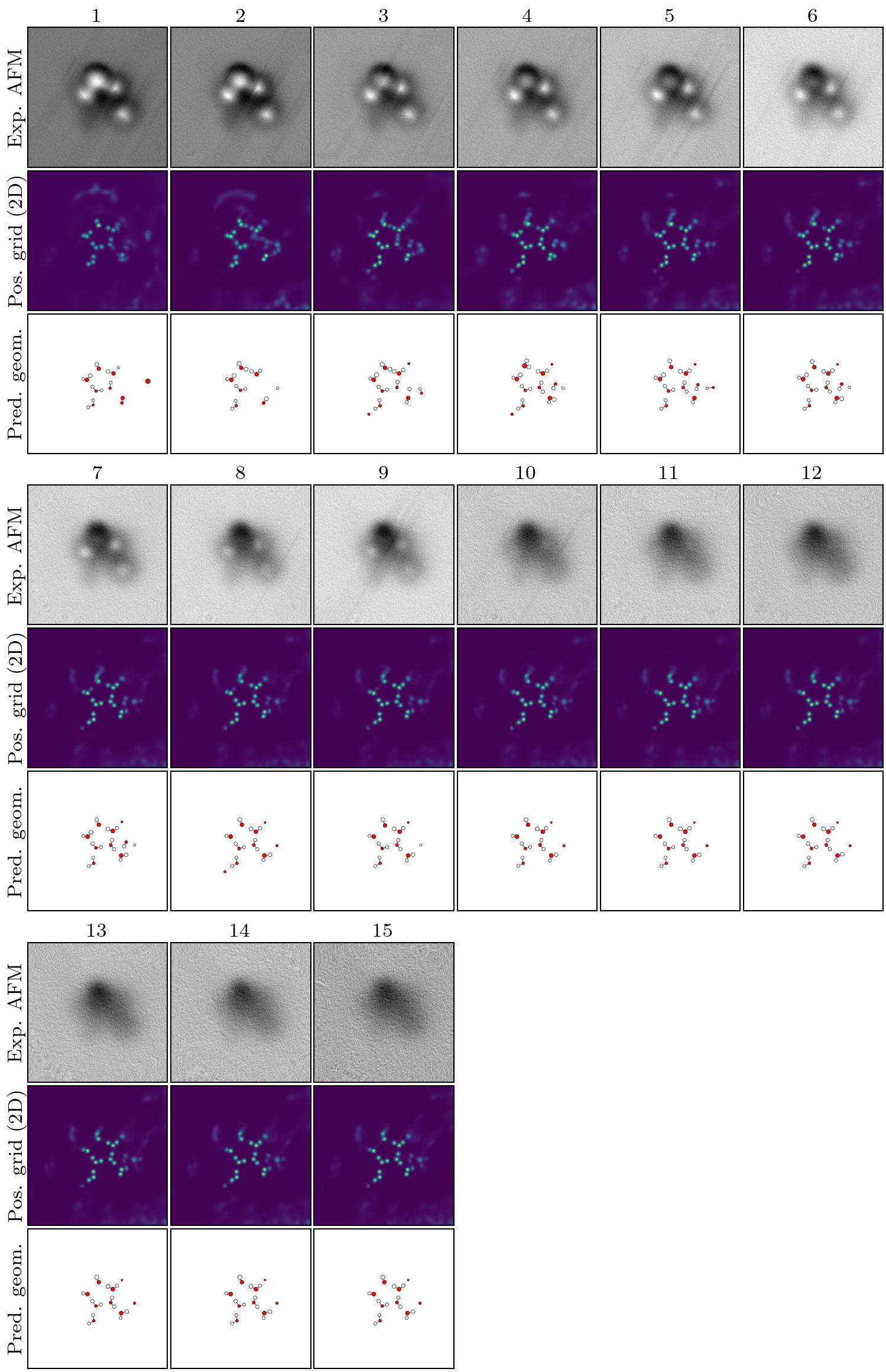}
	\caption{Atom position and graph predictions on experiment B as functions of the number of z-slices included in the AFM stack. Each of the predictions in columns 1-15 get as input the AFM image in that column and all the previous columns. In the first row are the AFM images, in the second row the prediction of the position prediction model (reduced from 3D to 2D by taking an average over the z dimension), and in the third row the prediction of the graph construction model.}
	\label{fig:pred_z}
\end{figure}

\begin{figure}
	\centering
	\includegraphics[width=125mm]{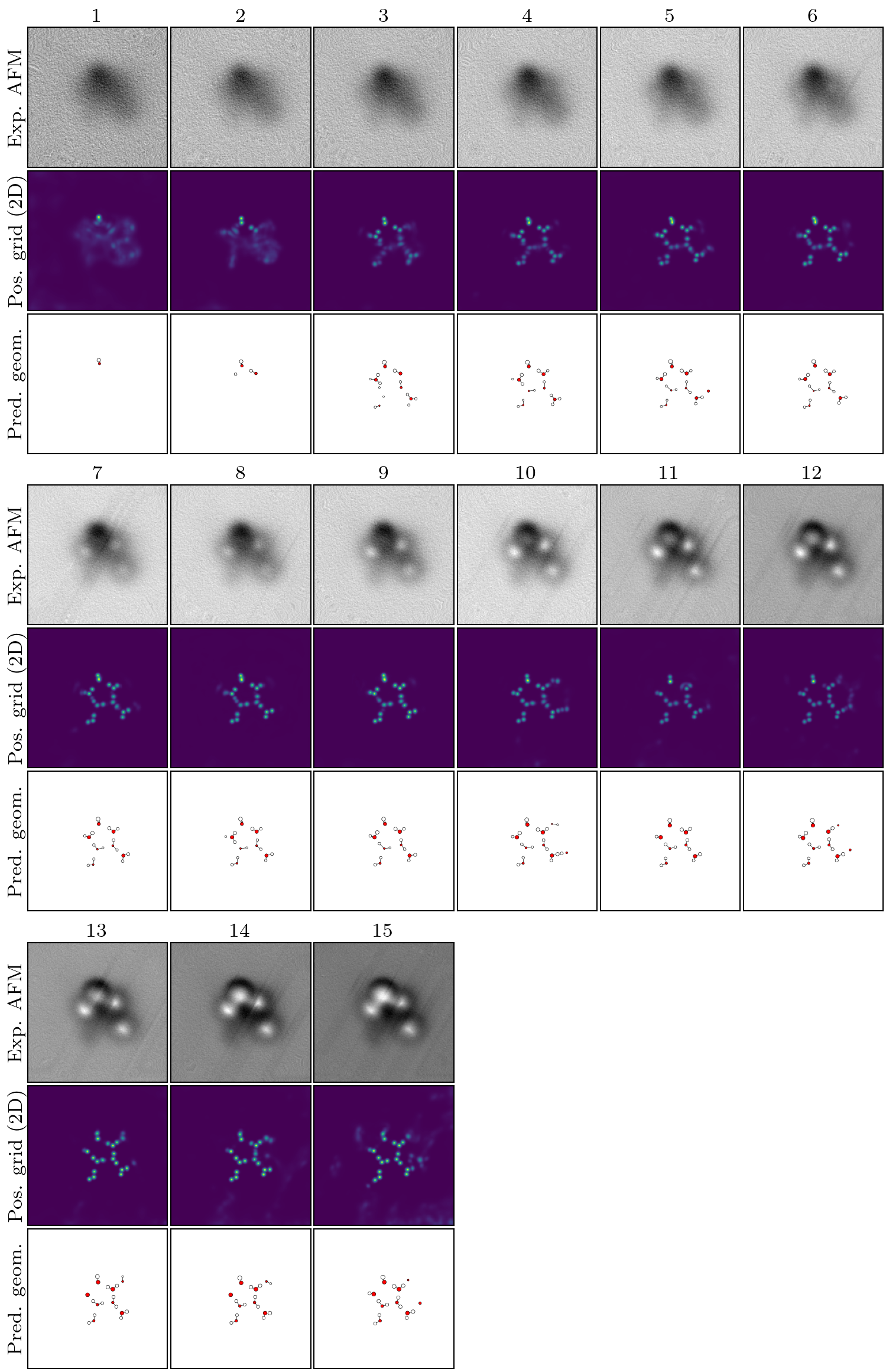}
	\caption{Same as Fig.~\ref{fig:pred_z} except with the order of the progression reversed.}
	\label{fig:pred_zr}
\end{figure}

Previous works that utilized CNNs for AFM image analysis always used a fixed number of AFM images in the input image stack, usually 10 images \cite{Alldritt2020, Oinonen2021, Oinonen2022, Tang2022, carracedo-cosme2022_molecular, carracedo-cosme2023}. Here we have modified the model to work with a variable stack size, which makes the model applicable to experiments that did not consider having a large enough vertical scan range as a requirement. Indeed, this is also the case for the experiments presented here, where the covered range in the vertical direction ranges between $0.7 \dots$ 1.4 \AA{} (see the full image stacks in Fig.~\ref{fig:afm_full}). In the main text we present the prediction using the full image stack for each experiment, but this raises the question if we could also get a good prediction with a smaller stack size, even going down to just a single image.

We start by considering the prediction accuracy on the simulation test set where we can perform a quantitative analysis. In Fig.~\ref{fig:loss_z}A is shown the loss for the position prediction model (CNN) as a function of the stack size. The stack size ranges between 1 and 15 images, starting from the closest approach and extending away from the sample. We can observe that the loss is clearly the highest for the single-image stack and decays quickly with increasing stack size. The loss saturates to a constant value at a stack size of roughly $n_z = 10$, suggesting that increasing the scan range to beyond $\sim \SI{1}{\angstrom}$ in the vertical direction does not bring any further benefit. Considering that many AFM experiments only aim to get a single sharp image of the sample, we also consider how the loss changes for a single image input as a function of the distance from the sample in Fig.~\ref{fig:loss_z}B. Here, the loss is the smallest at the closest distance, but the sharpest increase in happens at around $\Delta z = \SI{0.5}{\angstrom}$ offset from the closest distance, and the loss reaches a roughly constant value at $\Delta z \sim \SI{1.0}{\angstrom}$. This suggests, perhaps unsurprisingly, that for a single image the the closest range is the most informative. However, the curve also does not have a large gradient at $\Delta z = \SI{0.0}{\angstrom}$, which suggests that going any closer than the closest distance used in the training set may not bring any benefit. Although, it should be noted that the AFM simulations in this work use the Lennard-Jones potential for the Pauli interaction, and the result could be different with, for example, the full-density based model that provides a more accurate approximation of the Pauli repulsion \cite{ellner2019}.

Next, we consider how the prediction changes with increasing stack size on the experimental AFM images. We use the experiment B here as a case study, since it has the biggest stack size and none of the images had to be interpolated. We first consider the case where we start at the closest distance image and then stack more images on top, as shown in Fig.~\ref{fig:pred_z}. Already at the single-image prediction we get most of the geometry right, but a couple of the atoms are missing and there is an oxygen molecule in the place of one of the water molecules. The full structure is predicted at stack size 4, but there are some additional spurious atoms also predicted. The prediction finally becomes stable at stack size 10 and remains unchanged after that. This result is in line with the result of the analysis of the losses on the simulation test set. We also consider the reverse case where the image stack starts at the farthest distance and is grown towards the closer distances, which better corresponds to a situation in an experiment where the distance to the sample is closed gradually. The result is shown in Fig.~\ref{fig:pred_zr} where we can observe that the prediction for just one or two images only identifies a couple of the top atoms. However, the prediction improves very quickly, and the full structure is predicted at only five images in the stack. The prediction remains relatively stable when approaching closer, but we can also observe the prediction degrading at some of the closer images with an addition of false atom into the geometry. We cannot give a certain explanation for this behaviour, but we can observe that the closer images contain some tip artefacts stripes and the model may be trying to match these by placing false atoms at a deeper position that results in shadows in the image at those positions. Another factor may be the issue with the Lennard-Jones potential mentioned above, so that the training data may not represent the Pauli interaction at close range correctly.

We point out that the full geometry prediction at the first five images in Fig.~\ref{fig:pred_zr} is quite remarkable if one considers how seemingly little information the first five images contain by visual inspection. This seems almost impossible, but problem here is very constrained by the fact that the samples consist of only water molecules. This sufficiently limits the possible configurations so that the model can make a correct prediction from very little information. We do not expect this level of predictive power to generalize to models that make predictions in less constrained spaces, such as the varied kind of small organic molecules that have been typically imaged with high-resolution AFM.

\begin{figure}[h]
\centering
\begin{minipage}{.85\textwidth} 
\begin{algorithm}[H]
{ \fontsize{11}{12} \selectfont 
\caption{Generating candidate hydrogen bond networks.}
\begin{algorithmic}[1]
\For{$\text{n} = 1$ to $N_{\text{candidates}}$}
    \State Read cluster coordinates $\text{S}$.
    \State Select a random water molecule $\text{mol}_{\text{start}}$ in $\text{S}$
    \State \parbox[t]{.8\linewidth}{Reorder molecules in $\text{S}$ radially from $\text{mol}_{\text{start}}$, getting $\text{S}_{\text{radial}}$.\strut}
    \For{Each molecule in $\text{S}_{\text{radial}}$}
        \State \parbox[t]{.8\linewidth}{Assign probabilistically the molecule to either type "h" or "v", based on Figure \ref{fig:water_distribution} histograms.\strut}
        \If{Molecule is "h"}
            \State \parbox[t]{.8\linewidth}{Create bonds with two neighbors, if available with molecules already visited.\strut}
        \ElsIf{Molecule is "v"}
            \State \parbox[t]{.8\linewidth}{Create one bond with a neighbor, if available with molecules already visited, and direct the other towards the surface.\strut}
        \EndIf
        \If{No neighbors available for bonding}
            \If{Molecule is on the cluster edge}
                \State \parbox[t]{.8\linewidth}{Direct both hydrogens vertically.\strut}
            \Else
                \State \parbox[t]{.8\linewidth}{Randomly orient the molecule. \strut}
            \EndIf
        \EndIf
    \EndFor
\EndFor
\end{algorithmic}
} 
\end{algorithm}
\end{minipage}
\caption{Algorithm for generating a hydrogen bonds framework.}
\label{fig:hydrogen_bonds}
\end{figure}

\begin{figure}[h]
	\centering
	\includegraphics[width=150mm]{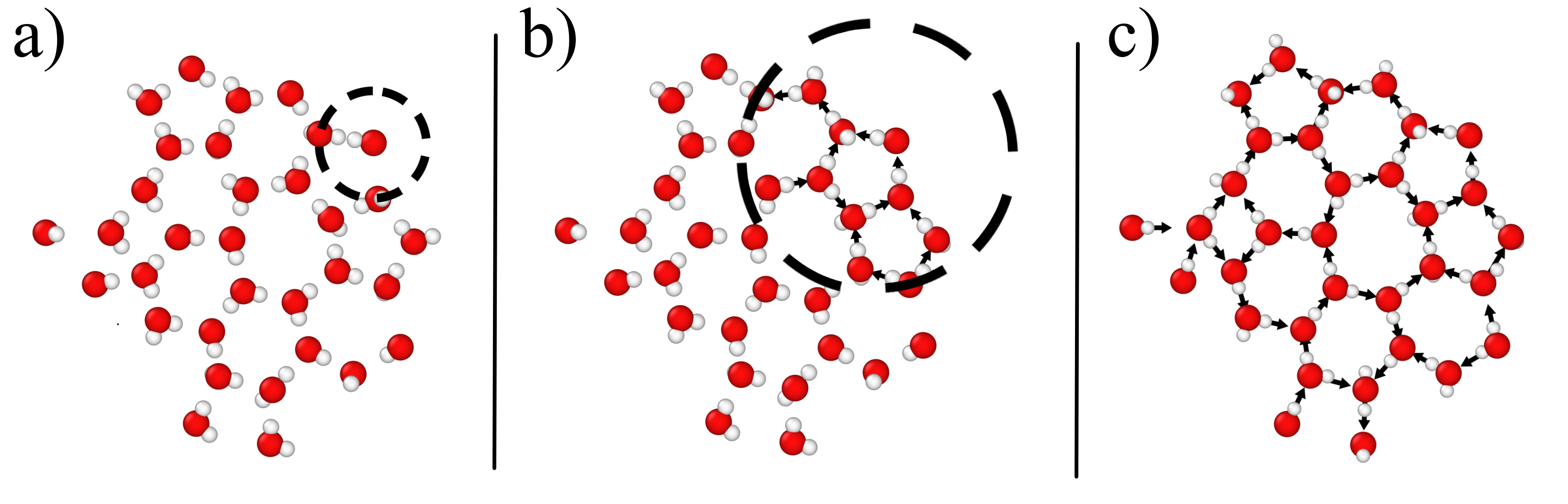}
	\caption{Example visualization of the algorithm in Fig.~\ref{fig:hydrogen_bonds}. In (a), one molecule of the cluster is randomly chosen to start the process. In (b), snapshot of an intermediate step, where newly formed bonds are indicated by the arrows. In (c), the final hydrogen bonds arrangement.}
	\label{fig:hydrogen_bonds_example}
\end{figure}

\clearpage

\section{Additional results}


\providecommand{\latin}[1]{#1}
\makeatletter
\providecommand{\doi}
  {\begingroup\let\do\@makeother\dospecials
  \catcode`\{=1 \catcode`\}=2 \doi@aux}
\providecommand{\doi@aux}[1]{\endgroup\texttt{#1}}
\makeatother
\providecommand*\mcitethebibliography{\thebibliography}
\csname @ifundefined\endcsname{endmcitethebibliography}
  {\let\endmcitethebibliography\endthebibliography}{}

\end{document}